\documentclass[journal]{./src/theme/vgtc/vgtc}                % final (journal style)

\ifpdf%                                % if we use pdflatex
  \pdfoutput=1\relax                   % create PDFs from pdfLaTeX
  \usepackage{graphicx}                % allow us to embed graphics files
  \DeclareGraphicsExtensions{.pdf,.png,.jpg,.jpeg} % for pdflatex we expect .pdf, .png, or .jpg files
\else%                                 % else we use pure latex
  \usepackage{graphicx}                % allow us to embed graphics files
  \DeclareGraphicsExtensions{.eps}     % for pure latex we expect eps files
\fi%

\graphicspath{{./src/figures/}{./img/}{./src/theme/vgtc/}{./}} % where to search for the images

\usepackage{microtype}                 % use micro-typography (slightly more compact, better to read)
\PassOptionsToPackage{warn}{textcomp}  % to address font issues with \textrightarrow
\usepackage{textcomp}                  % use better special symbols
\usepackage{mathptmx}                  % use matching math font
\usepackage{times}                     % we use Times as the main font
\usepackage{cite}                      % needed to automatically sort the references
\usepackage{tabu}                      % only used for the table example
\usepackage{booktabs}                  % only used for the table example

\usepackage[svgnames]{xcolor}%
\usepackage{dashbox}%

\onlineid{}
\vgtccategory{Research}
\vgtcpapertype{system}

\usepackage{comment,graphicx,amsmath,epsfig,listings,color,float}
\usepackage[utf8]{inputenc}
\usepackage{enumitem}
\usepackage{multicol}
\usepackage{tikz}
\usepackage{tikzscale}
\usepackage{lipsum}
\usepackage{ifthen}
\usepackage{multirow}

\usepackage{pgfplots}
\usepackage{filecontents}
\usepackage[binary-units=true]{siunitx}

\usetikzlibrary{colorbrewer,shapes,arrows,arrows.meta,calc,fit,positioning,patterns,decorations.markings,decorations.pathreplacing,backgrounds,tikzmark}

\usepackage{src/utils/changebarnofloat}
\usepackage{src/utils/circledsteps}
\usepackage{src/utils/wordbox}
\usepackage{src/utils/llosyms}
\usepackage{src/utils/operatorsymbols}

\definecolor{color-black}{RGB}{0,0,0}
\definecolor{color-grey}{RGB}{100,100,100}

\definecolor{color-set}{RGB}{229,196,148}
\definecolor{color-script}{RGB}{102,194,165}
\definecolor{color-sql}{RGB}{102,194,165}
\definecolor{color-fetch}{RGB}{141,160,203}
\definecolor{color-load}{RGB}{231,138,195}
\definecolor{color-input}{RGB}{252,141,98}
\definecolor{color-visualize}{RGB}{102,194,165}

\definecolor{color-combine}{RGB}{188,128,189}
\definecolor{color-hash}{RGB}{179,222,105}
\definecolor{color-sort}{RGB}{190,186,218}
\definecolor{color-sort2}{RGB}{204,235,197}
\definecolor{color-partition}{RGB}{141,211,199}
\definecolor{color-ordagg}{RGB}{128,177,211}
\definecolor{color-ordagg2}{RGB}{179,222,105}
\definecolor{color-window}{RGB}{253,180,98}
\definecolor{color-scan}{RGB}{170,170,170}
\definecolor{color-merge}{RGB}{251,128,114}

\definecolor{color-hash1}{RGB}{204,235,197}
\definecolor{color-hash2}{RGB}{179,222,105}

\definecolor{color-gradient1-1}{RGB}{128,177,211}
\definecolor{color-gradient1-2}{RGB}{141,211,199}
\definecolor{color-gradient1-3}{RGB}{204,235,197}
\definecolor{color-gradient1-4}{RGB}{179,222,105}

\definecolor{color-vanilla-join}{RGB}{253,180,98}
\definecolor{color-vanilla-b}{RGB}{251,128,114}
\definecolor{color-vanilla-group}{RGB}{255,237,111}

\definecolor{color-llosym-pipeline}{RGB}{190,186,218}
\definecolor{color-llosym-buffer}{RGB}{141,211,199}

\definecolor{color-traces-scan}{RGB}{217,95,2}
\definecolor{color-traces-hash}{RGB}{166,118,29}
\definecolor{color-traces-ordagg}{RGB}{230,171,2}
\definecolor{color-traces-sort}{RGB}{27,158,119}
\definecolor{color-traces-window}{RGB}{102,166,30}
\definecolor{color-traces-partition}{RGB}{117,112,179}
\definecolor{color-traces-combine}{RGB}{231,41,138}

\title{DashQL -- Complete Analysis Workflows with SQL}
\author{Andr\'{e} Kohn, Dominik Moritz, Thomas Neumann}
\shortauthortitle{DashQL -- SQL for Interactive Data Analytics}
\authorfooter{
\item
 Andr\'{e} Kohn and Thomas Neumann are with Technical University of Munich. E-mails: \{andre.kohn, thomas.neumann\}@in.tum.de.
\item
 Dominik Moritz is with the Carnegie Mellon University. E-mail: domoritz@cmu.edu.
}

\abstract{
We present DashQL, a language that describes complete analysis workflows in self-contained scripts.
DashQL combines SQL, the grammar of relational database systems, with a grammar of graphics in a grammar of analytics.
It supports preparing and visualizing arbitrarily complex SQL statements in a single coherent language.
The proximity to SQL facilitates holistic optimizations of analysis workflows covering data input, encoding, transformations, and visualizations.
These optimizations use model and query metadata for visualization-driven aggregation, remote predicate pushdown, and adaptive materialization.
We introduce the DashQL language as an extension of SQL and describe the efficient and interactive processing of text-based analysis workflows.
}

\keywords{Information visualization, systems, declarative specification}

\CCScatlist{ % not used in journal version
 \CCScat{K.6.1}{Management of Computing and Information Systems}%
{Project and People Management}{Life Cycle};
 \CCScat{K.7.m}{The Computing Profession}{Miscellaneous}{Ethics}
}

\teaser{
  \centering
  \input{src/tikz/pipeline}
  \vspace{0.1cm}
  \caption{
    Example of a DashQL script describing a complete analysis workflow.
    The script loads site activity data, resolves country names via ISO 3166 country codes, aggregates daily page views, and visualizes the data using an area chart and a table.
    DashQL statements unify user interactions, data loading, and the analysis in a single language.
  }
  \label{f:pipeline}
}

\vgtcinsertpkg

\usepackage[capitalize,noabbrev,nameinlink]{cleveref}

\begin{document}

\maketitle

\section{Introduction}
\label{sec:introduction}

Interactive Data Analysis has evolved as an umbrella term for diverse research around approachable, inspirational, explanatory, and efficient data processing.
Decades of prior work in these areas have assembled a comprehensive toolbox, guiding users on their paths towards valuable data insights.
A common principle among these tools has been the unification of graphics and database interactions, creating a gap towards database query languages like SQL.
Pioneering systems like Polaris (Tableau), for example, shield users from database specifics by pairing a graphic taxonomy with an own table algebra~\cite{polaris}.
This table algebra is lowered to SQL transparently which abstracts from subtle differences between SQL dialects and allows supporting various database systems through a single interface.

In the meantime, however, SQL has become the de-facto standard for data transforms at all scales, ranging from embedded systems (e.g., DuckDB~\cite{duckdb}, SQLite~\cite{sqlite}) to large data warehouses (e.g., Snowflake~\cite{snowflake}, F1~\cite{f1}, Procella~\cite{procella}, Presto~\cite{presto}, Redshift~\cite{redshift}, Azure Synapse~\cite{synapse}, CockroachDB~\cite{cockroachdb}, Hive~\cite{hive}).
Today's database abstractions therefore should match SQL in its expressivity or otherwise turn into an explicit translation layer that data analysts might have to work around.
This is particularly pronounced for advanced SQL functionality such as nested subqueries, non-inner joins, window aggregates or grouping sets that are often omitted as early victims during generalization.
A lack of these features can render today's tools insufficient for analysts that use SQL as their mental model for database interactions.

Additionally, database abstractions prevent holistic optimizations of analysis workflows.
The optimization of SQL queries is a well studied problem but is usually unaware of how data is ingested and how the results are consumed~\cite{reviewinteractivevis}.
Data analysts therefore propagate information back into the database, for example, by optimizing requests for a following visualization.
This is not only error prone but exposes relational optimizations to the user.
Database abstractions also obfuscate capabilities of the underlying database.
It is not uncommon in today's analysis tools to prepare and cache volatile data to optimize the repeated and interactive query evaluation during exploration~\cite{tableauresponse}.
However, this turns out to be a pitfall as it nullifies common database optimizations like projection and selection pushdown.
Structured file formats like Parquet allow reading data partially based on the query columns and filters.
A tool that lacks these query-driven optimizations might therefore be \emph{slower} if it loads \emph{unnecessary} data for a workflow.

We expand the vision of Wu et al.~\cite{dvms} and propose a language for a Data Visualization Management System (DVMS) that embeds data retrieval, loading, and visualization into SQL.
We call this SQL dialect DashQL and explain how a single coherent language model can drive interactive analysis workflows.
\Cref{f:pipeline} shows the first example of a DashQL script that visualizes grouped timeseries data using an area chart and table.
In the figure, the input script on the left is translated to a graph of tasks that drives the parallel evaluation of statements.
The right side of the figure hints at the visual output of the script as an interactive dashboard including an input field at the top of the screen, followed by the two visualizations.

The contribution of this paper is twofold.
We first introduce the language grammar and statement semantics in \Cref{sec:lang} and outline how a SQL dialect facilitates interactive exploration, scalable dashboards, and workflow development.
We then describe the efficient evaluation of DashQL workflows in \Cref{sec:implementation} and present holistic optimizations that use metadata for remote predicate pushdown and adaptive materialization.
\Cref{sec:implementation} also introduces AM4, an optimization in DashQL that accelerates visualizations with time series data.
We demonstrate DashQL examples throughout the paper and author an interactive analysis workflow step-by-step in \Cref{sec:evalexplore}.
We measure the performance of the holistic optimization AM4 in \Cref{sec:evalam4}.
We close with a discussion of related work in \Cref{sec:relatedwork} and a summary of the paper in \Cref{sec:summary}.

\section{Grammar of Analytics}
\label{sec:lang}

DashQL unifies the predominant grammar of relational~\cite{codd2002relational} database systems, SQL, with a grammar of graphics~\cite{wilkinson2012grammar} into a grammar of analytics.
This section introduces the DashQL language and its role in an analytics system.
We first list the grammar rules of DashQL and describe the semantics of every new statement.
Afterward, we present three advantages of driving analysis workflows with a coherent analysis language.

\subsection{SQL Extension}

\begin{figure}
    \centering
    \includegraphics[width=1.0\linewidth]{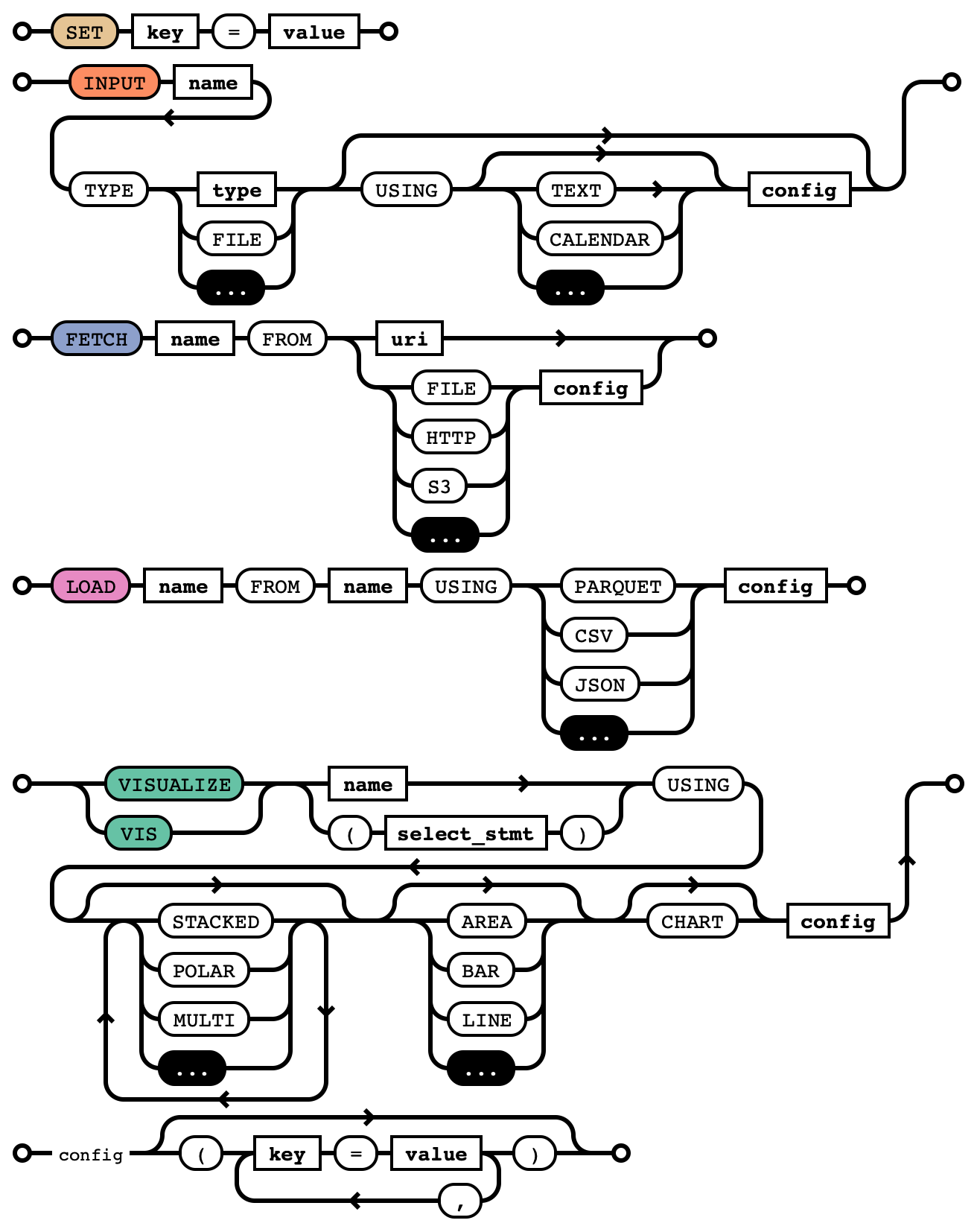}
    \vspace{0.1cm}
    \caption{Grammar rules of the five DashQL statements \texttt{SET}, \texttt{INPUT}, \texttt{FETCH}, \texttt{LOAD} and \texttt{VISUALIZE} that are combined with the statement rules of PostgreSQL.}
    \label{fig:grammarrails}
\end{figure}

DashQL introduces the five statements \wordbox[fill=color-set]{\texttt{SET}}, \wordbox[fill=color-input]{\texttt{INPUT}}, \wordbox[fill=color-fetch]{\texttt{FETCH}}, \wordbox[fill=color-load]{\texttt{LOAD}} and \wordbox[fill=color-visualize]{\texttt{VISUALIZE}} to the SQL language.
Together, they extend SQL just enough to specify where data is located, how it can be loaded and how it can be visualized for users.
This allows DashQL to describe complete analysis workflows in self-contained scripts, while preserving the expressiveness of arbitrary SQL queries.
The grammar rules of all statements are shown in \cref{fig:grammarrails} and are outlined in the following.

\wordbox[fill=color-set]{\texttt{SET}} is a utility statement that defines global script properties as individual key-value pairs.
This allows modifying script evaluation settings or provide script metadata such as titles, descriptions or versions.

\wordbox[fill=color-input]{\texttt{INPUT}} declares values that are provided to the script at runtime.
For example, an \texttt{INPUT} statement with identifier \texttt{x} and value type \texttt{FILE} presents an input control to users that opens a file picker dialog when clicked.
The provided file is then exposed to the remainder of the script through the identifier \texttt{x}.
\texttt{INPUT} may further be followed by an explicit component type and configuration options, matching additional settings like default values.

\wordbox[fill=color-fetch]{\texttt{FETCH}} accompanies \texttt{INPUT} as the second statement that declares raw data for DashQL scripts.
In its simplest form, the rule \texttt{FETCH name FROM uri} specifies a raw data source as a single URI.
The value \texttt{"https://a/b.parquet"}, for example, declares a remote file that will be loaded using \texttt{HTTPS}.
If a simple URI is not sufficient, the statement may alternatively be written with the explicit keyword \texttt{HTTPS} followed by fine-granular settings such as the method type or request headers.
Similarly, the \texttt{FETCH} statement allows supporting additional source types such as AWS S3, following the same syntax.

The fourth statement \wordbox[fill=color-load]{\texttt{LOAD}} defines how raw data can be loaded into the database.
We deliberately separate the fetching of opaque data and the extraction of relations since the boundary between these two can be fuzzy and depends on the capabilities of the underlying database.
DuckDB, for example, can partially scan remote Parquet files over \texttt{HTTPS} using a dedicated table function and a virtual file system abstraction.
This collapses both statements into following SQL statements, emphasizing the decoupled nature of declarations in a script and their efficient execution.
Other databases without these capabilities might need to execute either one or both tasks explicitly upfront.
Similar to \texttt{FETCH}, the \texttt{LOAD} statement can also be extended with additional keywords to introduce new data formats to the language.

\begin{figure*}
    \input{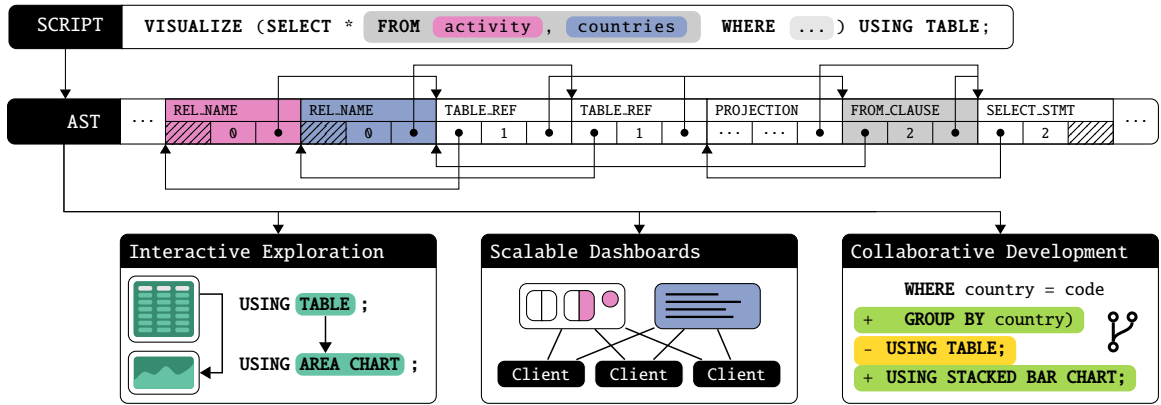}
    \caption{
        DashQL scripts as a driver for data analysis workflows.
        AST nodes store the location in the input text, the node type, the attribute key, the index of the parent node, and either a raw value or a span of children nodes.
        Script-based analysis workflows allow for interactive exploration, scalable dashboards, and a collaborative workflow development.
    }
    \label{f:overview}
\end{figure*}

The last statement \wordbox[fill=color-visualize]{\texttt{VISUALIZE}} displays data using charts and tables.
Creating visualizations is an iterative process that benefits from short round-trip times between ideas and their realizations.
DashQL offers approachable and fast exploration by combining a simple and short syntax with a fallback to a full grammar of graphics for refinements.
After all, visualizations are created in tandem with SQL statements which already provide useful information such as the attribute order of SQL projections or data types.

For example, users might want to display timeseries data with a time attribute \texttt{t}, and a value attribute \texttt{v} backed by a SQL query such as \texttt{CREATE TABLE a AS SELECT t, v \placeholder}.
Creating the first visualization for this table can be as simple as \texttt{VISUALIZE a USING LINE}.
Without further information, DashQL assumes that \texttt{t} and \texttt{v} were deliberately provided as first and second attributes referring to \texttt{x} and \texttt{y} values of the line chart.
Alternatively, SQL column aliases can be used in anticipation of ambiguities to name \texttt{x} and \texttt{y} explicitly, as in \texttt{SELECT v AS y, t AS x \placeholder}, .
Later iterations may then add a third attribute in the SQL projection list followed by \texttt{VISUALIZE a USING MULTI LINE}, in which case the new attribute would map to the line color.
This example shows that the interplay with traditional SQL statements may provide enough information to simplify the syntax significantly through carefully chosen defaults.

This simplified syntax enables rapid prototyping but may not suffice for advanced analysis reports.
The grammar therefore supports an alternative grammar rule with \texttt{VISUALIZE a USING (\placeholder)} that describes visualizations using raw Vega-Lite specifications.
In fact, the \emph{simple} form is lowered down to Vega-Lite internally, enabling automatic rewrites as explicit specification when refinements are needed.
In alignment with SQL, the Vega-Lite attributes are case insensitive in DashQL and JSON objects are replaced with nested key-value pair lists enclosed by round brackets.
The systematic generation and completion of Vega-Lite specifications is described in \Cref{sec:complementingvega}.

\subsection{Driving Analysis Workflows}

An extended SQL dialect offers an opportunity to describe entire analysis workflows in self-contained scripts.
These scripts become the single source of truth for an analysis system and can drive features such as interactive exploration, scalable dashboards, and collaborative workflow development.
\Cref{f:overview} illustrates these features based on a single visualization statement that displays the results of a join as a table.

\subsubsection{Interactive Exploration}

DashQL demystifies system internals by replacing a multitude of configuration knobs with guided textual editing.
DashQL strikes a balance between flexibility and intuition by providing short and long versions of the different grammar rules.
This flexibility allows users to start the exploration with short statements and later refine the workflow by manually adjusting inferred properties.
This simplifies the exploration as the syntactical differences between statements stay small.

The example in \Cref{f:overview} expresses the intent to display data as a table by writing \texttt{VISUALIZE \placeholder\ USING TABLE}.
The short syntax allows altering the visualization quickly.
For example, a user can replace the keyword \texttt{TABLE} with \texttt{AREA CHART} to change the visualization type and add the keyword \texttt{STACKED} to group areas based on an additional attribute.
Once the correct chart type is found, a user can adjust fine-granular configuration options such as colors and labels through explicit Vega-Lite settings.
These rewrites can either be done manually by changing the script text or by modifying a previously rendered chart.
The result is an interactive loop where partially evaluated DashQL workflows guide users through following refinements.

Additionally, the analysis workflows are interactive themselves through the \texttt{INPUT} statement.
These statements parameterize worklows explicitly by exposing variables to viewers.
This allows embedding arbitrary complex SQL queries into the workflow and steer them through input controls.
A popular alternative to \texttt{INPUT} statements is to derive raw SQL text from input values through text interpolation.
This is flexible, but complicates the semantic analysis of workflows as it gives up crucial information about statement dependencies, types, and the exact usage of parameters.
It also requires a preprocessing step to generate the actual script text which does not align well with the continuous and iterative re-evaluation of workflows.

DashQL distinguishes between the analysts authoring workflows and the viewer that consume the workflows's output.
Viewers are not exposed to the language but instead only see the results from statements as opaque analysis dashboards.
Authors, in contrast, see the language and visual output side-by-side and benefit from semantic information in the script editor.
Interactive exploration in DashQL is therefore emphasized differently for these two user groups as authors benefit from frictionless feedback loops for textual changes while viewers require efficient re-evaluations after changing input values.

\subsubsection{Scalable Dashboards}

DashQL simplifies the sharing of analysis dashboards.
A workflow is a single self-contained script text and can be treated as such for the distribution to multiple users.
This decouples the workflow description from the evaluating analysis tool, similarly to SQL being the common denominator between relational database systems.
Sharing a data analysis workflow is \textit{cheap} since there is no dependency on specific service resources except for the workflow's input data.
The price for serving this data is often \textit{lower} than maintaining computing resources for traditional server-based analytics tools, more so with scalable cloud storage services and large content delivery networks.

We introduce the language DashQL alongside a reference implementation that is powered by DuckDB-Wasm, an efficient WebAssembly version of the analytical database DuckDB~\cite{duckdb} for the web.
It evaluates entire analytical workflows ad-hoc in the browser, presenting a cost-efficient and interactive solution without dedicated analytics servers.
The lack of a dedicated server increases the horizontal scalability of the system at the cost of higher bandwidth requirements for the viewers.
According to Vogelsgesang et al, shared analysis workflows on smaller datasets are not uncommon today~\cite{getreal}.
They state, that only approximately 600 out of 62 thousand workbooks uploaded to the service Tableau Public contain more than a million tuples.
All other workflows fall into the range of browser-manageable data sizes, eliminating the need for dedicated computing resources in the cloud.

DashQL also supports workflows that process larger datasets but reduce the data size quickly based on user input.
For example, if a workflow processes event data of a logging service, the entire dataset for all users might easily exceed petabytes of records.
But if the workflow itself analyzes events of a specific user over a fixed period, the datasets can get sufficiently small.
\Cref{sec:holisticopt} introduces holistic optimizations that optimize the amount of loaded data based on SQL queries in the workflow.
Yet, the language DashQL is not limited to small datasets.
It instead offers an opportunity to dynamically combine client and server-side implementations to optimize for scalability and interactivity wherever possible and fall back to traditional server-side processing when needed.

\begin{figure*}
    \input{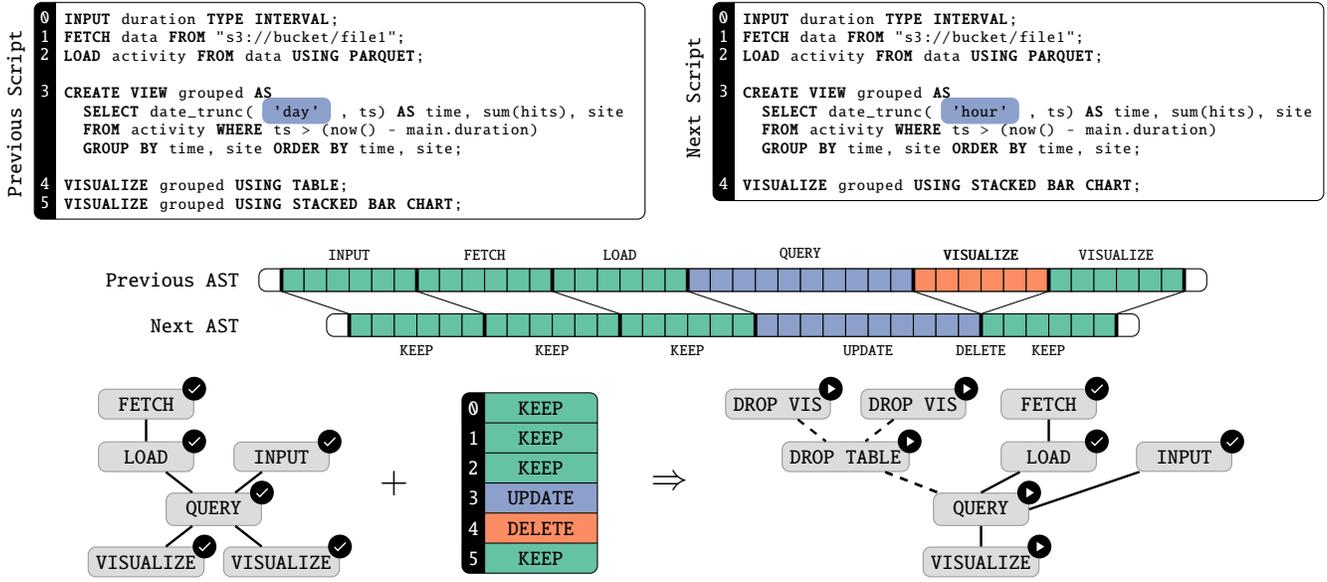}
    \vspace{0.2cm}
    \caption{
        Example of a task graph that is derived from a previous task graph and an AST-based script difference.
        The two scripts visualize grouped timeseries data and differ in a deleted statement and the grouping granularity.
        The AST colors equal statements in \wordbox[fill=color-visualize]{green}, changes in \wordbox[fill=color-fetch]{blue} and deletions in \wordbox[fill=color-input]{orange}.}
    \label{f:programdiff}
\end{figure*}

\subsubsection{Collaborative Development}

DashQL also simplifies collaborative development of analysis workflows.
Text-based version control systems like \emph{Git} dominate the distributed software development today.
Since DashQL workflows are self-contained scripts, they can be developed as part of a versioned development process.
Users can fork DashQL workflows and contribute changes back through simple textual updates.
This process is facilitated by the concise grammar of SQL that keeps the textual differences small.
\Cref{f:overview} demonstrates this versioning by adding a grouping clause and a changed chart type in the example statement.
The patch tracks the new grouping by the country attribute as well as the visualization as stacked bar chart in the same script.
DashQL workflows can therefore be created, updated, forked and discussed in environments that have already proven their effectiveness in collaborative development.
\section{Implementation}
\label{sec:implementation}

In this section, we outline the implementation of a DashQL powered analysis tool.
We first describe the efficient AST encoding that we use as textual language model for analysis workflows.
This model allows the runtime to update only the parts of the execution state that have changed instead of full re-evaluations.
We then introduce the concept of tasks and show how adaptive task graphs can be maintained using fast difference computations.
We discuss the extensibility of DashQL and the use of query metadata to simplify declarative visualizations for fast exploration.
And finally, we present two examples of holistic optimizations that are accelerating the coupled workflow components.

\subsection{AST Format}

DashQL translates many user interactions into modifications of the associated script text.
This positions the underlying text model as fundamental component of the entire system.
Our implementation is therefore built around a fast syntactical analysis, backed by an efficient representation of the abstract syntax tree (AST).
The parser extends the SQL grammar rules of PostgreSQL and allocates compact AST nodes into a single, bump-allocated memory buffer. 
This accelerates parsing and increases the cache efficiency of any following operations such as tree traversals.
An AST node is exactly \SI{20}{\byte} large and stores the location in the input text, the node type, the attribute key and the index of the parent node.
It also stores either a raw integer value or a span of children nodes in the same buffer.
The text location associates each node with the substring matched by its grammar rule, enabling partial rewrites of individual statements.
The AST further acts as an auxiliary data structure and references string literals in the original script text instead of copying them.
Children of an AST node are further stored in sorted order based on the attribute key, accelerating key lookups and recursive comparisons.

\Cref{f:overview} illustrates the AST encoding of an example statement that visualizes an inline SQL query joining two base relations.
The AST presents two nodes of type \texttt{REL\_NAME} that match the table names \texttt{A} and \texttt{B}.
Nodes are created by the parser following a post-order traversal of matched grammar rules which emits children before their parents.
This eliminates additional serialization steps since nodes can be written to the buffer while parsing.
In the example, the relation names are matched as table references and then form the two children of a from clause.
The output of the syntactic analysis is a program description representing statements as offsets of root nodes in this AST buffer.
This representation is not only cache efficient, but also simplifies the crossing of system boundaries as the consecutive memory buffer and fixed size nodes simplify the communication between system components and languages.

\subsection{From AST to Task}

The actionable units of our system are called tasks.
Tasks are derived from statements and form a graph based on the statement dependencies.
A query statement that references a load statement, for example, translates into a query task that scans the output of a load task.
These tasks are partially ordered and evaluate the entire script starting with data ingestion and ending with the visualization of derived tables.
New tasks are derived on every user interaction based on the difference between the AST and its predecessor.
This includes tasks to undo the effects of deleted statements, update the effects of modified statements and add the effects of new statements.

For example, if a SQL statement that created a table is deleted from the script, the system derives a task to undo the effects by dropping the table.
This mechanism is more abstract than the traditional transaction isolation of database systems as all tasks are maintaining a single workflow state that is updated with respect to changes in the script text and the user input.
\texttt{VISUALIZE} statements, for example, compile Vega-Lite specifications only once and delete or update the specification only when the statement changes.
\texttt{FETCH} statements that download data using HTTP will further cache the data until a script change invalidates the output.
The task graph drives the execution of an analysis workflow and serves as anchor for any operations on derived state.

\subsection{Adaptive Task Graphs}

The task graph of DashQL is adaptive as it reflects all continuous changes in the script and the user input.
We implement a variant of an algorithm known as \emph{Patience Diff} that is implemented in the version control systems \emph{GNU Bazaar} and \emph{Git}.
The algorithm derives task updates from the difference between two scripts and works as follows:

We first determine all unique statement mappings between a script and its predecessor.
Two statements are compared based on their ASTs instead of texts for whitespace insensitivity and support for incremental changes.
The similarity can be quantified by counting equal AST nodes in two simultaneous DFS traversals and weighting them by the distance to the AST root.
The tree traversals profit from the compact and cache efficient encoding of nodes into a single AST buffer.
Next, we compute the longest common subequence among the mapped statements and use them as anchors for the remaining assignments.
The remainder is then iterated in sequence and assigned to the most similar matches that haven't been assigned yet.
This identifies new and deleted statements and emits a similarity score for the rest.

Afterward, we determine the \emph{applicability} of all previous tasks.
A task is \emph{applicable} if it was derived from a statement that stayed the same, does not \emph{transitively} depend on an \emph{inapplicable} task, and is not followed by an \emph{inapplicable} task that successfully modified the own output.
The \emph{applicability} can be determined through a single DFS traversal with backwards propagation when encountering an \emph{inapplicable} task.
\emph{Applicable} tasks and their state are migrated and marked as completed while the effects of all other tasks are updated or undone.

\cref{f:programdiff} illustrates the entire process with an example of two scripts that analyze site activity data stored in AWS S3.
The first script starts with an \texttt{INPUT} statement that receives a time interval for the analysis.
It then downloads the data from an AWS S3 bucket using a \texttt{FETCH} statement and inserts the data into the database as Parquet file using \texttt{LOAD}.
The statements are followed by a traditional SQL query to filter the site activity in the input interval and compute aggregates grouped by days.
The final two statements visualize the result of this query as a table and a stacked bar chart.
The second script is almost identical to the first one except that the data is now grouped per hour instead of days and is no longer visualized as table.

AST buffers of both scripts are shown below with the node color indicating the statement differences.
The first three statements and the last are equal and therefore don't need to change.
The query statement differs in the string literal that is passed to the function \texttt{date\_trunc} and is marked as updated.
The first visualization statement is no longer present in the new script and is marked as deleted.

The figure also contains a task graph derived for the previous script.
It shows one task for every statement and a checkmark indicating that all of them were successfully executed.
This task graph is then combined with the computed statement mappings to derive a new set of tasks reflecting the changes between the scripts.
The table visualization was deleted, emitting a task called \texttt{DROP VIZ} to remove the table.
The query statement was updated and results in the task \texttt{DROP TABLE} to undo the effects of the SQL query.
However, this effect propagates since both visualizations depend on the table data.
We therefore also undo the effects of the second visualization and recreate it after executing the updated SQL statement.
The remaining tasks that fetch and load the remote data into the database and receive the input from the user are migrated and marked as completed.

This example demonstrates differences with the traditional script execution in relational database systems.
DashQL is defining entire analysis workflows, including external data, visualizations, and interactions with a user.
Scripts are therefore not evaluated independently but in the context of a preceding execution, rewarding awareness of existing state.

\subsection{Complementing Vega-Lite}
\label{sec:complementingvega}

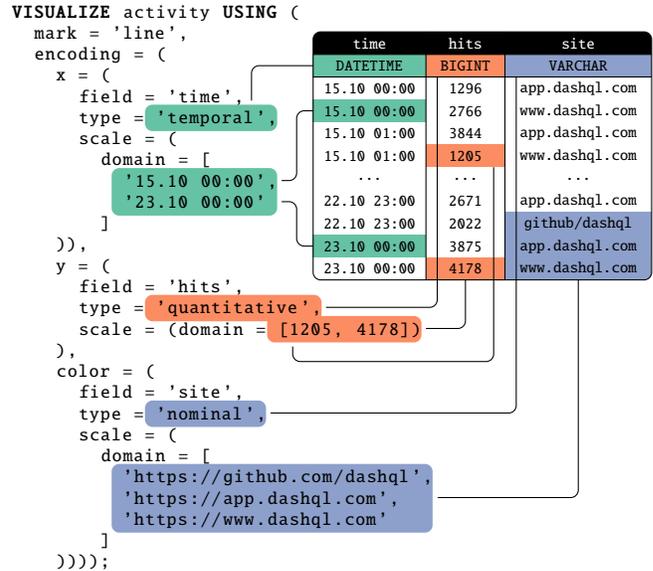
\begin{figure}[t]
    \definecolor{color-x}{RGB}{102,194,165}
\definecolor{color-y}{RGB}{252,141,98}
\definecolor{color-color}{RGB}{141,160,203}

\pgfmathsetmacro\cs{0.3}
\pgfmathsetmacro\rowh{0.3}
\pgfmathsetmacro\rowcount{11}
\pgfmathsetmacro\lastrow{10}

\tikzstyle{tbl-cell}=[anchor=center,font=\fontsize{6pt}{7pt}\selectfont\ttfamily]
\tikzstyle{tbl-idx}=[anchor=center,font=\fontsize{8pt}{9pt}\selectfont\ttfamily,color=white]
\tikzstyle{origin}=[draw, rounded corners=1mm]
\tikzstyle{textbox}=[inner ysep=5pt,inner xsep=4pt,yshift=2pt,rounded corners=1mm]

\lstdefinestyle{codegen}{emph={INPUT,AS,INTERVAL,VARCHAR,FETCH,LOAD,USING,CREATE,TABLE,VIEW,SELECT,FROM,WHERE,AND,VISUALIZE,LINE,PARQUET,CSV,HTTP,INTO,GROUPED,MULTI,CHART,VEGA},emphstyle={\bfseries}}
\lstset{
    style = codegen,
    numbers = none,
    escapechar = !,
    basicstyle=\fontsize{7}{8}\selectfont\ttfamily
}

\begin{tikzpicture}[remember picture, overlay]
  \def\column#1#2#3#4#5{
      \foreach \i in {0,...,#3} {
          \coordinate(#1;\i;nw) at ($(#2) + (0, -\i * #5)$);
          \coordinate(#1;\i;ne) at ($(#1;\i;nw) + (#4, 0)$);
          \coordinate(#1;\i;se) at ($(#1;\i;nw) + (#4, -1.0 * #5)$);
          \coordinate(#1;\i;sw) at ($(#1;\i;nw) + (0, -1.0 * #5)$);
          \coordinate(#1;\i;s) at ($(#1;\i;nw) + (0.5 * #4, -1.0 * #5)$);
          \coordinate(#1;\i;w) at ($(#1;\i;nw) + (0, -0.5 * #5)$);
          \coordinate(#1;\i;e) at ($(#1;\i;nw) + (#4, -0.5 * #5)$);
          \coordinate(#1;\i;center) at ($(#1;\i;nw) + (0.5 * #4, -0.5 * #5)$);
      }
  }
  \coordinate(tbl-anchor) at ($(pic cs:lst3-mark-1) + (6*\cs,0.3*\cs)$);
  \column{tbl-0}{tbl-anchor}{\rowcount}{5*\cs}{\rowh};
  \column{tbl-1}{tbl-0;0;ne}{\rowcount}{3.5*\cs}{\rowh};
  \column{tbl-2}{tbl-1;0;ne}{\rowcount}{6.5*\cs}{\rowh};

  \path[fill=color-x] (tbl-0;1;nw) -- (tbl-0;1;ne) -- (tbl-0;1;se) -- (tbl-0;1;sw)  -- cycle;
  \path[fill=color-y] (tbl-1;1;nw) -- (tbl-1;1;ne) -- (tbl-1;1;se) -- (tbl-1;1;sw)  -- cycle;
  \path[fill=color-color] (tbl-2;1;nw) -- (tbl-2;1;ne) -- (tbl-2;1;se) -- (tbl-2;1;sw)  -- cycle;
  \path[fill=black] (tbl-0;0;nw) -- (tbl-0;0;ne) -- (tbl-0;0;se) -- (tbl-0;0;sw) {[rounded corners=1mm] -- cycle};
  \path[fill=black] (tbl-1;0;nw) -- (tbl-1;0;ne) -- (tbl-1;0;se) -- (tbl-1;0;sw) -- cycle;
  \path[fill=black] (tbl-2;0;nw) {[rounded corners=1mm] -- (tbl-2;0;ne)} -- (tbl-2;0;se) -- (tbl-2;0;sw) -- cycle;

  \path[fill=color-color] (tbl-2;10;nw) -- (tbl-2;10;ne) {[rounded corners=1mm] -- (tbl-2;10;se)} -- (tbl-2;10;sw) -- cycle;
  \path[fill=color-color] (tbl-2;9;nw) -- (tbl-2;9;ne) -- (tbl-2;9;se) -- (tbl-2;9;sw) -- cycle;
  \path[fill=color-color] (tbl-2;8;nw) -- (tbl-2;8;ne) -- (tbl-2;8;se) -- (tbl-2;8;sw) -- cycle;
  \path[fill=color-x] (tbl-0;9;nw) -- (tbl-0;9;ne) -- (tbl-0;9;se) -- (tbl-0;9;sw) -- cycle;
  \path[fill=color-x] (tbl-0;3;nw) -- (tbl-0;3;ne) -- (tbl-0;3;se) -- (tbl-0;3;sw) -- cycle;
  \path[fill=color-y] (tbl-1;5;nw) -- (tbl-1;5;ne) -- (tbl-1;5;se) -- (tbl-1;5;sw) -- cycle;
  \path[fill=color-y] (tbl-1;10;nw) -- (tbl-1;10;ne) -- (tbl-1;10;se) -- (tbl-1;10;sw) -- cycle;
  
  \path[draw, rounded corners=1mm] (tbl-0;0;nw) -- (tbl-2;0;ne) -- (tbl-2;\lastrow;se) -- (tbl-0;\lastrow;sw) -- cycle;
  \path[draw] (tbl-0;0;ne) -- (tbl-0;\lastrow;se);
  \path[draw] (tbl-1;0;ne) -- (tbl-1;\lastrow;se);
  \path[draw] (tbl-0;0;sw) -- (tbl-2;0;se);
  \path[draw] (tbl-0;1;sw) -- (tbl-2;1;se);

	% \path[fill=black] (tbl-0;0;nw) -- (tbl-0;0;ne) -- (tbl-0;0;se) -- (tbl-0;0;sw) {[rounded corners=1mm] -- cycle};
	% \foreach \i in {1,...,4} {
	% 	\path[fill=black] (tbl-0;\i;nw) -- (tbl-0;\i;ne) -- (tbl-0;\i;se) -- (tbl-0;\i;sw) -- cycle;
	% }
	% \path[fill=black] (tbl-0;5;nw) -- (tbl-0;5;ne) -- (tbl-0;5;se) {[rounded corners=1mm] -- (tbl-0;5;sw)} -- cycle;

  \node[tbl-cell,color=white] at (tbl-0;0;center) {time};
  \node[tbl-cell] at (tbl-0;1;center) {DATETIME};
  \node[tbl-cell] at (tbl-0;2;center) {15.10 00:00};
  \node[tbl-cell] at (tbl-0;3;center) {15.10 00:00};
  \node[tbl-cell] at (tbl-0;4;center) {15.10 01:00};
  \node[tbl-cell] at (tbl-0;5;center) {15.10 01:00};
  \node[tbl-cell] at (tbl-0;6;center) {...};
  \node[tbl-cell] at (tbl-0;7;center) {22.10 23:00};
  \node[tbl-cell] at (tbl-0;8;center) {22.10 23:00};
  \node[tbl-cell] at (tbl-0;9;center) {23.10 00:00};
  \node[tbl-cell] at (tbl-0;10;center) {23.10 00:00};

  \node[tbl-cell,color=white] at (tbl-1;0;center) {hits};
  \node[tbl-cell] at (tbl-1;1;center) {BIGINT};
  \node[tbl-cell] at (tbl-1;2;center) {1296};
  \node[tbl-cell] at (tbl-1;3;center) {2766};
  \node[tbl-cell] at (tbl-1;4;center) {3844};
  \node[tbl-cell] at (tbl-1;5;center) {1205};
  \node[tbl-cell] at (tbl-1;6;center) {...};
  \node[tbl-cell] at (tbl-1;7;center) {2671};
  \node[tbl-cell] at (tbl-1;8;center) {2022};
  \node[tbl-cell] at (tbl-1;9;center) {3875};
  \node[tbl-cell] at (tbl-1;10;center) {4178};

  \node[tbl-cell,color=white] at (tbl-2;0;center) {site};
  \node[tbl-cell] at (tbl-2;1;center) {VARCHAR};
  \node[tbl-cell] at (tbl-2;2;center) {app.dashql.com};
  \node[tbl-cell] at (tbl-2;3;center) {www.dashql.com};
  \node[tbl-cell] at (tbl-2;4;center) {app.dashql.com};
  \node[tbl-cell] at (tbl-2;5;center) {www.dashql.com};
  \node[tbl-cell] at (tbl-2;6;center) {...};
  \node[tbl-cell] at (tbl-2;7;center) {app.dashql.com};
  \node[tbl-cell] at (tbl-2;8;center) {github/dashql};
  \node[tbl-cell] at (tbl-2;9;center) {app.dashql.com};
  \node[tbl-cell] at (tbl-2;10;center) {www.dashql.com};

  \coordinate (xdomain-0) at ($(pic cs:lst3-x-domain-0-1) + (0.2, 0.08)$);
  \coordinate (xdomain-1) at ($(pic cs:lst3-x-domain-0-1) + (0.4, 0.08)$);
  \coordinate (xdomain-2) at ($(pic cs:lst3-x-domain-1-1) + (0.2, 0.08)$);
  \coordinate (xdomain-3) at ($(pic cs:lst3-x-domain-1-1) + (0.4, 0.08)$);
  \coordinate (xdomain-4) at ($(tbl-0;3;w)$);
  \coordinate (xdomain-5) at ($(tbl-0;9;w)$);
  \path[origin] (xdomain-0) -- (xdomain-1) |- (xdomain-4);
  \path[origin] (xdomain-2) -- (xdomain-3) |- (xdomain-5);

  \coordinate (temporal-0) at ($(pic cs:lst3-x-type-1) + (-0.2, 0.3)$);
  \coordinate (temporal-1) at ($(tbl-0;1;w)$);
  \path[origin] (temporal-0) |- (temporal-1);

  \coordinate (quant-0) at ($(pic cs:lst3-y-type-1) + (0.2, 0.08)$);
  \coordinate (quant-1) at ($(tbl-1;1;sw) + (0.15, 0.0)$);
  \path[origin] (quant-0) -| (quant-1);

  \coordinate (nominal-0) at ($(pic cs:lst3-color-type-1) + (0.2, 0.08)$);
  \coordinate (nominal-1) at ($(tbl-2;1;sw) + (0.15, 0.0)$);
  \path[origin] (nominal-0) -| (nominal-1);

  \coordinate (ydomain-0) at ($(pic cs:lst3-y-scale-domain-1) + (0.2, 0.08)$);
  \coordinate (ydomain-1) at (tbl-1;10;s);
  \coordinate (ydomain-2) at ($(pic cs:lst3-y-scale-domain-0) + (0.2, -0.16)$);
  \coordinate (ydomain-3) at ($(tbl-1;5;se) - (0.15, 0)$);
  \path[origin] (ydomain-0) -| (ydomain-1);
  \path[origin] (ydomain-2) -- ++(0, -0.2) -| (ydomain-3);

  \coordinate (scale-1) at ($(pic cs:lst3-color-scale-range-1-1) + (0.65, 0.08)$);
  \coordinate (scale-3) at ($(tbl-2;10;s) + (0, 0)$);
  \path[origin] (scale-1) -| (scale-3);

  \node[textbox,fill=color-x,fit={(pic cs:lst3-x-type-0) (pic cs:lst3-x-type-1)}] {};
  \node[textbox,fill=color-x,fit={(pic cs:lst3-x-domain-0-0) (pic cs:lst3-x-domain-1-1)}] {};
  \node[textbox,fill=color-y,fit={(pic cs:lst3-y-type-0) (pic cs:lst3-y-type-1)}] {};
  \node[textbox,fill=color-y,fit={(pic cs:lst3-y-scale-domain-0) (pic cs:lst3-y-scale-domain-1)}] {};
  \node[textbox,fill=color-color,fit={(pic cs:lst3-color-type-0) (pic cs:lst3-color-type-1)}] {};
  \node[textbox,fill=color-color,fit={(pic cs:lst3-color-scale-range-0-0) (pic cs:lst3-color-scale-range-0-1) (pic cs:lst3-color-scale-range-1-1) (pic cs:lst3-color-scale-range-2-1)}] {};
\end{tikzpicture}

\begin{lstlisting}
VISUALIZE activity USING MULTI LINE CHART;

VISUALIZE activity USING (
  mark = !\tikzmark{lst3-mark-0}!'line'!\tikzmark{lst3-mark-1}!,
  encoding = (
    x = (
      field = !\tikzmark{lst3-x-field-0}!'time'!\tikzmark{lst3-x-field-1}!,
      type = !\tikzmark{lst3-x-type-0}!'temporal'!\tikzmark{lst3-x-type-1}!,
      scale = (
        domain = [
          !\tikzmark{lst3-x-domain-0-0}!'15.10 00:00'!\tikzmark{lst3-x-domain-0-1}!,
          !\tikzmark{lst3-x-domain-1-0}!'23.10 00:00'!\tikzmark{lst3-x-domain-1-1}!
        ]
    )),
    y = (
      field = !\tikzmark{lst3-y-field-0}!'hits'!\tikzmark{lst3-y-field-1}!,
      type = !\tikzmark{lst3-y-type-0}!'quantitative'!\tikzmark{lst3-y-type-1}!,
      scale = (domain = !\tikzmark{lst3-y-scale-domain-0}![1205, 4178]!\tikzmark{lst3-y-scale-domain-1}!)
    ),
    color = (
      field = !\tikzmark{lst3-color-field-0}!'site'!\tikzmark{lst3-color-field-1}!,
      type = !\tikzmark{lst3-color-type-0}!'nominal'!\tikzmark{lst3-color-type-1}!,
      scale = (
        domain = [
          !\tikzmark{lst3-color-scale-range-0-0}!'https://github.com/dashql'!\tikzmark{lst3-color-scale-range-0-1}!,
          !\tikzmark{lst3-color-scale-range-1-0}!'https://app.dashql.com'!\tikzmark{lst3-color-scale-range-1-1}!,
          !\tikzmark{lst3-color-scale-range-2-0}!'https://www.dashql.com'!\tikzmark{lst3-color-scale-range-2-1}!
        ]
    ))));
\end{lstlisting}
    \caption{
        Two \texttt{VISUALIZE} statements that produce the same time series line chart, showing website hits of multiple websites.
        DashQL generates Vega-Lite specifications based on the table schema and statistics.}
    \label{f:vegacompletion}
\end{figure}

Vega-Lite offers a grammar to describe an expressive range of charts in declarative JSON specifications. 
The \texttt{VISUALIZE} statement of DashQL supports Vega-Lite specifications as nested key-value pair lists in SQL.
\texttt{VISUALIZE} does not need to embed its own grammar of graphics and users already familiar with Vega-Lite don't have to learn a new language.

Vega-Lite specifications are self-contained and describe visualizations without the context of an existing data model.
In DashQL, visualizations are always backed by SQL queries which offer an opportunity to auto-complete parts of a specification.
This reduces the pressure on Vega-Lite and pushes costly data introspection into the database system.
An example for this are encoding types and scale domains.
We know the data types of all involved attributes based on the SQL metadata which enables robust defaults, for example, when selecting between \emph{quantitative}, \emph{ordinal}, and \emph{nominal} encoding types.
Additionally, we can determine a value domain or range efficiently upfront using SQL queries.

DashQL further provides simplified \texttt{VISUALIZE} statements that can be written in tandem with SQL queries.
This follows the observation, that explicit defaults can guide the writing of SQL queries with respect to a subsequent visualization.
For example, users can express a preferred field assignment through attribute aliases.
A projection like \texttt{SELECT time AS x, hits AS y, site AS color \placeholder} implictly provides the fields for a visualization that can be reduced to \texttt{VISUALIZE \placeholder\ USING MULTI LINE CHART}.
These mappings can still be overruled by explicit settings but enable short statement variants for fast exploration.
Even without matching aliases, users can fall back to listing attributes in a certain order.
By default, DashQL assigns the first three attributes to the encoding channels \texttt{x}, \texttt{y}, \texttt{color}, resulting in the same output.

\cref{f:vegacompletion} lists two \texttt{VISUALIZE} statements.
The first one presents a simplified syntax that instructs DashQL to draw a chart with multiple lines.
The second one describes the same output using an embedded Vega-Lite specification.
Internally, DashQL uses the table on the right to derive the specification of the second statement for the first one.
The table has the three columns \texttt{time}, \texttt{hits} and \texttt{site} with the data types \texttt{DATETIME}, \texttt{BIGINT} and \texttt{VARCHAR}.
By default, a chart with multiple lines requires encoding declarations for \texttt{x}-values, \texttt{y}-values and the line \texttt{color}.
Without further hints, DashQL assigns the columns in-order, using the time attribute for \texttt{x}, the number of hits for \texttt{y}, and the site name as \texttt{color}.
The encoding types are derived based on the encoded column and the column data types.
An \texttt{x}-encoding backed by a \texttt{DATETIME} attribute is \emph{temporal}, by default.
The other encoding types are \emph{quantitative} for \texttt{y}-values of type \texttt{BIGINT} and \emph{nominal} for \texttt{color}-values of type \texttt{VARCHAR}.
DashQL then resolves the domain for each of the scales in the encodings.
The domains of temporal and quantitative scales are computed using minimum and maximum aggregates and yield the interval between the 15th and 23nd of August for the time attribute and the value range between 1205 and 4178 for the hit count.
The domain of the nominal scale is then resolved by querying distinct site values, emitting the three websites of the DashQL project.

In summary, Vega-Lite provides a robust grammar for declarative visualizations in DashQL.
We further extend the capabilities of Vega-Lite by completing specifications based on the contextual query metadata.
This accelerates the data exploration without losing the flexibility of a full specification whenever it is needed.

\subsection{Language Extensions}

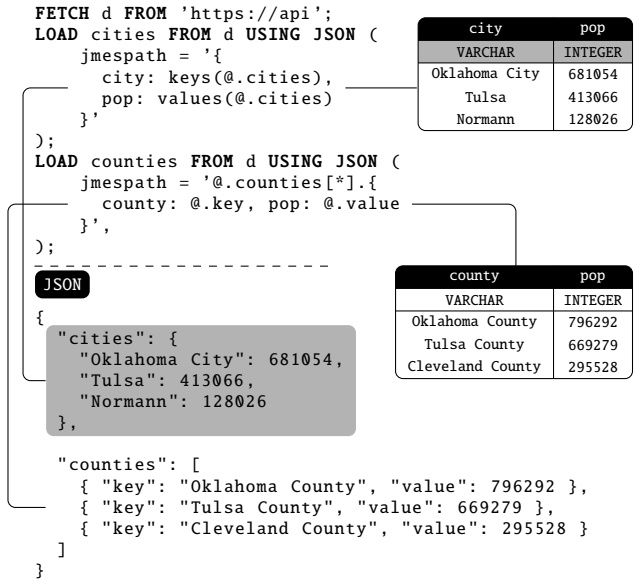
\begin{figure}[t]
    \lstdefinestyle{codegen}{emph={INPUT,AS,INTERVAL,VARCHAR,FETCH,LOAD,USING,CREATE,TABLE,VIEW,SELECT,FROM,WHERE,AND,VISUALIZE,LINE,PARQUET,CSV,HTTP,INTO,GROUPED,MULTI,CHART,VEGA,JSON},emphstyle={\bfseries}}
\definecolor{color-json0}{RGB}{179,179,179}
\definecolor{color-json1}{RGB}{255,255,255}
\definecolor{color-rel0}{RGB}{179,179,179}
\definecolor{color-rel1}{RGB}{255,255,255}
\definecolor{color-expr}{RGB}{230,230, 230}

\pgfmathsetmacro\cs{0.3}
\pgfmathsetmacro\rowh{0.3}
\pgfmathsetmacro\rowcount{4}
\pgfmathsetmacro\lastrow{4}

\tikzstyle{tbl-cell}=[anchor=center,font=\fontsize{6pt}{7pt}\selectfont\ttfamily]
\tikzstyle{tbl-idx}=[anchor=center,font=\fontsize{8pt}{9pt}\selectfont\ttfamily,color=white]

\tikzstyle{jsonlabel}=[anchor=south west,font=\fontsize{7pt}{7pt}\selectfont\ttfamily,color=white,fill=black,rounded corners=1mm]
\tikzstyle{textbox}=[inner ysep=5pt,inner xsep=4pt,yshift=2pt,rounded corners=1mm]

\begin{tikzpicture}[remember picture, overlay]
  \def\column#1#2#3#4#5{
      \foreach \i in {0,...,#3} {
          \coordinate(#1;\i;nw) at ($(#2) + (0, -\i * #5)$);
          \coordinate(#1;\i;ne) at ($(#1;\i;nw) + (#4, 0)$);
          \coordinate(#1;\i;se) at ($(#1;\i;nw) + (#4, -1.0 * #5)$);
          \coordinate(#1;\i;sw) at ($(#1;\i;nw) + (0, -1.0 * #5)$);
          \coordinate(#1;\i;s) at ($(#1;\i;nw) + (0.5 * #4, -1.0 * #5)$);
          \coordinate(#1;\i;w) at ($(#1;\i;nw) + (0, -0.5 * #5)$);
          \coordinate(#1;\i;e) at ($(#1;\i;nw) + (#4, -0.5 * #5)$);
          \coordinate(#1;\i;center) at ($(#1;\i;nw) + (0.5 * #4, -0.5 * #5)$);
      }
  }

  % TEXT BOXES
  \node[textbox,fill=color-json0,fit={(pic cs:lst9-1) (pic cs:lst9-2) (pic cs:lst9-9)}] (rel0-json) {};
  \node[textbox,fit={(pic cs:lst9-5) (pic cs:lst9-6) (pic cs:lst9-10) (pic cs:lst9-11)}] (rel1-load) {};
  \node[textbox,fill=color-json1,fit={(pic cs:lst9-3) (pic cs:lst9-4) (pic cs:lst9-14)}] (rel1-json) {};
  \node[textbox,fit={(pic cs:lst9-7) (pic cs:lst9-13) (pic cs:lst9-15)}] (rel0-load) {};

  % TABLE 1
  \coordinate(tbl-anchor) at ($(pic cs:lst9-20) + (17*\cs,1*\cs)$);
  \column{tbl0-0}{tbl-anchor}{\rowcount}{6*\cs}{\rowh};
  \column{tbl0-1}{tbl0-0;0;ne}{\rowcount}{3.5*\cs}{\rowh};

  \path[fill=white, rounded corners=1mm] (tbl0-0;0;nw) -- (tbl0-1;0;ne) -- (tbl0-1;\lastrow;se) -- (tbl0-0;\lastrow;sw) -- cycle;
  \path[fill=color-rel0] (tbl0-0;1;nw) -- (tbl0-0;1;ne) -- (tbl0-0;1;se) -- (tbl0-0;1;sw)  -- cycle;
  \path[fill=color-rel0] (tbl0-1;1;nw) -- (tbl0-1;1;ne) -- (tbl0-1;1;se) -- (tbl0-1;1;sw)  -- cycle;
  \path[draw, rounded corners=1mm] (tbl0-0;0;nw) -- (tbl0-1;0;ne) -- (tbl0-1;\lastrow;se) -- (tbl0-0;\lastrow;sw) -- cycle;
  \path[fill=black] (tbl0-0;0;nw) -- (tbl0-0;0;ne) -- (tbl0-0;0;se) -- (tbl0-0;0;sw) {[rounded corners=1mm] -- cycle};
  \path[fill=black] (tbl0-1;0;nw) {[rounded corners=1mm] -- (tbl0-1;0;ne)} -- (tbl0-1;0;se) -- (tbl0-1;0;sw) -- cycle;
  \path[draw] (tbl0-0;0;ne) -- (tbl0-0;\lastrow;se);
  \path[draw] (tbl0-0;0;sw) -- (tbl0-1;0;se);
  \path[draw] (tbl0-0;1;sw) -- (tbl0-1;1;se);

  \node[tbl-cell,color=white] at (tbl0-0;0;center) {city};
  \node[tbl-cell] at (tbl0-0;1;center) {VARCHAR};
  \node[tbl-cell] at (tbl0-0;2;center) {Oklahoma City};
  \node[tbl-cell] at (tbl0-0;3;center) {Tulsa};
  \node[tbl-cell] at (tbl0-0;4;center) {Normann};

  \node[tbl-cell,color=white] at (tbl0-1;0;center) {pop};
  \node[tbl-cell] at (tbl0-1;1;center) {INTEGER};
  \node[tbl-cell] at (tbl0-1;2;center) {681054};
  \node[tbl-cell] at (tbl0-1;3;center) {413066};
  \node[tbl-cell] at (tbl0-1;4;center) {128026};

  % TABLE 2
  \coordinate(tbl-anchor) at ($(pic cs:lst9-20) + (16*\cs,-10*\cs)$);
  \column{tbl1-0}{tbl-anchor}{\rowcount}{7*\cs}{\rowh};
  \column{tbl1-1}{tbl1-0;0;ne}{\rowcount}{3.5*\cs}{\rowh};

  \path[fill=white,rounded corners=1mm] (tbl1-0;0;nw) -- (tbl1-1;0;ne) -- (tbl1-1;\lastrow;se) -- (tbl1-0;\lastrow;sw) -- cycle;
  \path[fill=color-rel1] (tbl1-0;1;nw) -- (tbl1-0;1;ne) -- (tbl1-0;1;se) -- (tbl1-0;1;sw)  -- cycle;
  \path[fill=color-rel1] (tbl1-1;1;nw) -- (tbl1-1;1;ne) -- (tbl1-1;1;se) -- (tbl1-1;1;sw)  -- cycle;
  \path[draw,rounded corners=1mm] (tbl1-0;0;nw) -- (tbl1-1;0;ne) -- (tbl1-1;\lastrow;se) -- (tbl1-0;\lastrow;sw) -- cycle;
  \path[fill=black] (tbl1-0;0;nw) -- (tbl1-0;0;ne) -- (tbl1-0;0;se) -- (tbl1-0;0;sw) {[rounded corners=1mm] -- cycle};
  \path[fill=black] (tbl1-1;0;nw) {[rounded corners=1mm] -- (tbl1-1;0;ne)} -- (tbl1-1;0;se) -- (tbl1-1;0;sw) -- cycle;
  \path[draw] (tbl1-0;0;ne) -- (tbl1-0;\lastrow;se);
  \path[draw] (tbl1-0;0;sw) -- (tbl1-1;0;se);
  \path[draw] (tbl1-0;1;sw) -- (tbl1-1;1;se);

  \node[tbl-cell,color=white] at (tbl1-0;0;center) {county};
  \node[tbl-cell] at (tbl1-0;1;center) {VARCHAR};
  \node[tbl-cell] at (tbl1-0;2;center) {Oklahoma County};
  \node[tbl-cell] at (tbl1-0;3;center) {Tulsa County};
  \node[tbl-cell] at (tbl1-0;4;center) {Cleveland County};

  \node[tbl-cell,color=white] at (tbl1-1;0;center) {pop};
  \node[tbl-cell] at (tbl1-1;1;center) {INTEGER};
  \node[tbl-cell] at (tbl1-1;2;center) {796292};
  \node[tbl-cell] at (tbl1-1;3;center) {669279};
  \node[tbl-cell] at (tbl1-1;4;center) {295528};

  \path[draw,rounded corners=1mm] (rel0-json.west) --++ (-0.3, 0) |- (rel0-load.west);
  \path[draw,rounded corners=1mm] (rel1-json.west) --++ (-0.5, 0) |- (rel1-load.west);
  \path[draw,rounded corners=1mm] (rel0-load.east) --++ (0.95,0);
  \path[draw,rounded corners=1mm] (rel1-load.east) -| ($(tbl1-0;0;ne) + (-0.5,0)$);

  \coordinate (delim-anchor) at ($(pic cs:lst9-16) + (0, 0.75)$);
  \path[draw,dashed] (delim-anchor) --++ (4, 0);
  \node[jsonlabel, anchor=north west] at ($(delim-anchor) + (0, -0.05)$) {JSON};

\end{tikzpicture}

\lstset{
    style = codegen,
    numbers = none,
    escapechar = !,
    basicstyle=\fontsize{7}{8}\selectfont\ttfamily
}
\begin{lstlisting}[xleftmargin=0.4cm]
FETCH d FROM 'https://api';
!\tikzmark{lst9-20}!LOAD cities FROM d USING JSON (
    !\tikzmark{lst9-7}!jmespath = '{
      city: keys(@.cities),!\tikzmark{lst9-13}!
      pop: values(@.cities)
    }'!\tikzmark{lst9-15}!
);
!\tikzmark{lst9-21}!LOAD counties FROM d USING JSON (
    !\tikzmark{lst9-11}!jmespath = '@.counties[*].{!\tikzmark{lst9-5}!
      county: @.key, pop: @.value!\tikzmark{lst9-10}!
    }',!\tikzmark{lst9-6}!
);!\vspace{0.38cm}!

!\tikzmark{lst9-16}!{
  !\tikzmark{lst9-1}!"cities": {
    "Oklahoma City": 681054,!\tikzmark{lst9-9}!
    "Tulsa": 413066,
    "Normann": 128026
  },!\tikzmark{lst9-2}!

  !\tikzmark{lst9-3}!"counties": [
    { "key": "Oklahoma County", "value": 796292 },
    { "key": "Tulsa County", "value": 669279 },
    { "key": "Cleveland County", "value": 295528 }!\tikzmark{lst9-14}!
  ]!\tikzmark{lst9-4}!
}
\end{lstlisting}
    \caption{
        Two load statements that extract two relations from a single JSON document using JMESPath expressions.
        Both expressions extract populations in Oklahoma.
        The first expression emits the city data in column-major format, the second expression returns county data in row-major format.
    }
    \label{f:jmespath}
\end{figure}

The syntax of the DashQL statements \texttt{INPUT}, \texttt{FETCH}, \texttt{LOAD} and \texttt{VISUALIZE} end with optional settings provided as key-value pair lists.
This offers a mechanism to extend DashQL without modifying the grammar rules or the language model.
The settings translate to a generic dictionary that is passed to the derived tasks.
Custom task implementations can read this dictionary and enable extensions based on available keys.

Our reference implementation, for example, extends the loading of JSON data through JMESPath expressions.
By default, our embedded database DuckDB-Wasm can load a table from a JSON document in two formats.
Either in row-major format as top-level array of objects where each object contains all attributes of the relation or in column-major format as top-level object with members storing column arrays.
If a JSON document is not in either of those formats, it has to be transformed first.
The JSON task therefore checks for the key "jmespath" in the settings.
If it is present, the task evaluates the expression on the input data first before loading it into the database.

\Cref{f:jmespath} lists an example DashQL script, that loads two relations from a single JSON document that was returned from a remote HTTP API.
The documents stores population data of Oklahoma.
City populations are stored as a single object with city names as properties, whereas county populations are provided as an array of objects.
The first expression emits an object with the field \emph{city} storing an array of city names and the field \emph{pop} holding an array of population values.
The second expression returns the county object array with changed attribute names.
This example demonstrates the extensibility of the DashQL language through custom task implementations that can be configured through dynamic configuration options.

\subsection{Holistic Optimization}
\label{sec:holisticopt}

Data transformations can be expensive which makes their optimization indespensable for every data analysis workflow.
Query optimizers are therefore a vital component of every data processing system today and have a significant impact on overall execution times.
Research around query optimization is profound and has been expanded for decades.
Yet, databases are universal and face the difficult task to accelerate specific queries without losing the generality.
As a result, database systems rarely include external information during planning, leaving these non-trival problems to the applications.
DashQL unifies the data retrieval, transformations, and visualizations in the same language which presents an opportunity for holistic optimizations.

\begin{figure}
    \input{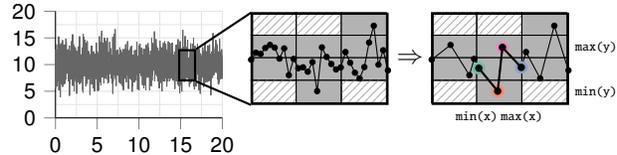}
    \caption{
        M4, a query for value-preserving time series aggregation, described by Jugel et al \cite{m4}.
        This version uses a CTE instead of a subquery with equal semantics.}
    \label{f:m4}
\end{figure}

\begin{figure}
    \definecolor{color-focus}{RGB}{252,141,98}
\definecolor{color-xmin}{RGB}{102,194,165}
\definecolor{color-xmax}{RGB}{141,160,203}
\definecolor{color-ymin}{RGB}{252,141,98}
\definecolor{color-ymax}{RGB}{231,138,195}
\tikzstyle{textbox}=[inner ysep=5pt,inner xsep=4pt,yshift=2pt,rounded corners=1mm]
\tikzstyle{am4-label}=[font=\fontsize{5pt}{6pt}\selectfont\ttfamily]
\tikzstyle{implies}=[font=\fontsize{10pt}{11pt}\selectfont\ttfamily,rounded corners=1mm,minimum width=1.2em,minimum height=1.2em]

\begin{tikzpicture}[
    %Environment Cfg.
    font=\bfseries\sffamily,
]
\begin{axis}[
    width=3.8cm,
    height=3cm,
    at={(0,0)},
    ymin=0,
    ymax=20,
    xmin=0,
    xmax=20,
    grid=both,
    minor tick num =1,
    minor tick style={draw=none},
    minor grid style={thin,color=black!10},
    major grid style={thin,color=black!10},
    ylabel={},
    xlabel={},
    tick align=outside,
    axis x line*=middle,
    axis y line*=none,
    xtick={0,5,...,20},
    ytick={0,5,...,20},
    x tick label style={
        /pgf/number format/assume math mode, font=\sf\scriptsize},
    y tick label style={
        /pgf/number format/assume math mode, font=\sf\scriptsize},
]
\addplot[color=black!60] table [x index=0,y index=1,col sep=comma] {./src/data/am4_noisy.csv};
\end{axis}

\newcommand\pixelgrid[6]{
    \foreach \i [evaluate=\i as \x using \i*#3/#5] in {0,...,#5} {
        \foreach \j [evaluate=\j as \y using \j*#4/#6] in {0,...,#6} {
            \coordinate(#1;\i;\j;nw) at ($#2 + (\x, -\y)$);
            \coordinate(#1;\i;\j;ne) at ($#2 + (\x + #3 / #5, -\y)$);
            \coordinate(#1;\i;\j;sw) at ($#2 + (\x, -\y - #4 / #6)$);
            \coordinate(#1;\i;\j;se) at ($#2 + (\x + #3 / #5, -\y - #4 / #6)$);
        }
    }
}
\tikzstyle{cell-set}=[draw,fill=black!30]
\tikzstyle{cell-clear}=[draw,pattern=north east lines,pattern color=black!30]

\pixelgrid{pxL}{(2.6,1.4)}{1.8}{1.2}{3}{4};
\newcommand\pxL[3] {
    \path[#3] (pxL;#1;#2;nw) -- (pxL;#1;#2;ne) -- (pxL;#1;#2;se) -- (pxL;#1;#2;sw) -- cycle;
}
\pxL{0}{0}{cell-clear};
\pxL{0}{1}{cell-set};
\pxL{0}{2}{cell-set};
\pxL{0}{3}{cell-clear};
\pxL{1}{0}{cell-clear};
\pxL{1}{1}{cell-set};
\pxL{1}{2}{cell-set};
\pxL{1}{3}{cell-set};
\pxL{2}{0}{cell-set};
\pxL{2}{1}{cell-set};
\pxL{2}{2}{cell-set};
\pxL{2}{3}{cell-clear};

\pixelgrid{pxR}{(5.0,1.4)}{1.8}{1.2}{3}{4};
\newcommand\pxR[3] {
    \path[#3] (pxR;#1;#2;nw) -- (pxR;#1;#2;ne) -- (pxR;#1;#2;se) -- (pxR;#1;#2;sw) -- cycle;
}
\pxR{0}{0}{cell-clear};
\pxR{0}{1}{cell-set};
\pxR{0}{2}{cell-set};
\pxR{0}{3}{cell-clear};
\pxR{1}{0}{cell-clear};
\pxR{1}{1}{cell-set};
\pxR{1}{2}{cell-set};
\pxR{1}{3}{cell-set};
\pxR{2}{0}{cell-set};
\pxR{2}{1}{cell-set};
\pxR{2}{2}{cell-set};
\pxR{2}{3}{cell-clear};

% \newcommand\pxLbox[6] {
%     \path[#6] ($(pxL;#1;#2;nw) + (0, -#4)$) -- ($(pxL;#1;#2;ne) + (0, -#4)$) -- ($(pxL;#1;#3;se) + (0, #5)$) -- ($(pxL;#1;#3;sw) + (0, #5)$) -- cycle;
% }
% \pxLbox{1}{1}{3}{0.1}{0.1}{draw=red};

\node[color=black,anchor=center, draw, minimum width=0.2cm, minimum height=0.4cm,thick] (boxS) at (1.75,0.7) {};
\node[anchor=north west, draw, minimum width=1.8cm, minimum height=1.2cm,thick] (boxL) at (pxL;0;0;nw) {};
\node[anchor=north west, draw, minimum width=1.8cm, minimum height=1.2cm,thick] (boxR) at (pxR;0;0;nw) {};
\path[color=black,draw,thick] (boxS.north east) -- (boxL.north west);
\path[color=black,draw,thick] (boxS.south east) -- (boxL.south west);
\node[implies,anchor=west] at ($(boxL.east) + (0, 0)$) {$\Rightarrow$};

\begin{axis}[
    clip mode=individual,
    anchor=north west,
    at={(2.6cm,1.4cm)},
    width=3.4cm,
    height=2.8cm,
    hide axis,
    enlarge x limits=false,
    enlarge y limits=0.2,
    xmin=15.0,
    xmax=15.29,
]
\addplot[color=black, mark=*, mark size=1pt] table [x index=0,y index=1,col sep=comma] {./src/data/am4_noisy.csv};

\end{axis}

\begin{axis}[
    clip mode=individual,
    anchor=north west,
    at={(5.0cm,1.4cm)},
    width=3.4cm,
    height=2.8cm,
    hide axis,
    enlarge x limits=false,
    enlarge y limits=0.2,
    xmin=15.0,
    xmax=15.29,
]
\addplot[color=black, mark=*, mark size=1pt] table [x index=0,y index=1,col sep=comma] {./src/data/am4_reduced.csv};
\addplot[color=color-xmin, mark=*, mark size=2pt, thick] coordinates {(15.1,9.008802093467313)};
\addplot[color=color-ymin, mark=*, mark size=2pt, thick] coordinates {(15.14,4.948690621388153)};
\addplot[color=color-ymax, mark=*, mark size=2pt, thick] coordinates {(15.15,12.689372914589587)};
\addplot[color=color-xmax, mark=*, mark size=2pt, thick] coordinates {(15.19,9.149424527312991)};
\addplot[color=black, mark=*, mark size=1pt, thick] table [x index=0,y index=1,col sep=comma] {./src/data/am4_focus.csv};

\end{axis}

\node[am4-label,anchor=north] at (pxR;1;3;sw) {min(x)};
\node[am4-label,anchor=north] at (pxR;1;3;se) {max(x)};
\node[am4-label,anchor=west] at ($(pxR;2;1;ne) + (0, -0.16)$) {max(y)};
\node[am4-label,anchor=west] at ($(pxR;2;3;ne) + (0, -0.15)$) {min(y)};

\end{tikzpicture}

\begin{tikzpicture}[remember picture, overlay]

\node[textbox,fill=color-xmin,fit={(pic cs:lst4-xmin-0) (pic cs:lst4-xmin-1)}] (xmin) {};
\node[textbox,fill=color-xmax,fit={(pic cs:lst4-xmax-0) (pic cs:lst4-xmax-1)}] (xmax) {};
\node[textbox,fill=color-ymin,fit={(pic cs:lst4-ymin-0) (pic cs:lst4-ymin-1)}] (ymin) {};
\node[textbox,fill=color-ymax,fit={(pic cs:lst4-ymax-0) (pic cs:lst4-ymax-1)}] (ymax) {};
    
\end{tikzpicture}

\vspace{-0.4cm}

\centering
\lstdefinestyle{codegen}{emph={INPUT,TYPE,AS,INTERVAL,VARCHAR,FETCH,LOAD,USING,CREATE,TABLE,VIEW,SELECT,FROM,WHERE,AND,VISUALIZE,LINE,PARQUET,CSV,HTTP,INTO,GROUP,BY},emphstyle={\bfseries}}
\lstset{
    style = codegen,
    numbers = none,
    escapechar = !,
    basicstyle=\fontsize{7}{10}\selectfont\ttfamily\linespread{0.8}
}
\begin{lstlisting}
SELECT !\tikzmark{lst4-xmin-0}!min(x), arg_min(y, x)!\tikzmark{lst4-xmin-1}!, !\tikzmark{lst4-xmax-0}!max(x), arg_max(y, x)!\tikzmark{lst4-xmax-1}!,
       !\tikzmark{lst4-ymin-0}!min(y), arg_min(x, y)!\tikzmark{lst4-ymin-1}!, !\tikzmark{lst4-ymax-0}!max(y), arg_max(x, y)!\tikzmark{lst4-ymax-1}!,
       round($width * (x - $lb) / ($ub - $lb)) AS bin,
FROM ( $user_data ) GROUP BY bin
\end{lstlisting}
    \caption{AM4, a more efficient version of M4 that provides value-preserving time series aggregation using a single scan and the aggregation functions \texttt{arg\_min} and \texttt{arg\_max}.}
    \label{f:am4}
\end{figure}

% \begin{figure*}
%     \input{src/tikz/distributed}
%     \vspace{0.3cm}
%     \caption{
%         Disaggregated computation with DashQL.
%         The script on the right downloads a Parquet file with stock data and joins it with an input CSV file.
%         The left side presents multiple ways to execute the script in a distributed setting.
%         \Circled{0} shows the traditional separation between client and server, \Circled{2} a fully local execution, \Circled{1} a hybrid mode in between.
%     }
%     \label{f:distributed}
% \end{figure*}

\subsubsection{Visualization-Driven Aggregation}

The first example for holistic optimization is the automatic aggregation of SQL results for \texttt{VISUALIZE} statements.
Jugel et al introduced the value-preserving aggregation M4 \cite{m4} to accelerate the visualization of time series data.
M4 follows the observation, that the amount of rendered data points in line charts can be limited by the number of visible pixels on the screen.
Instead of visualizing every single tuple of a time series, we can select a subset of the tuples based on the chart dimensions.
The authors group values by time bins and compute the four name-giving aggregates \texttt{min(x)}, \texttt{max(x)}, \texttt{min(y)}, and \texttt{max(y)} per bin.
The associated points span a bounding box around all tuples in a bin that intersects any pixels that should be colored for the line chart.
With DashQL, introducing M4 becomes an optimization that propagates the visualization context towards the backing SQL query.

\begin{figure*}
    \input{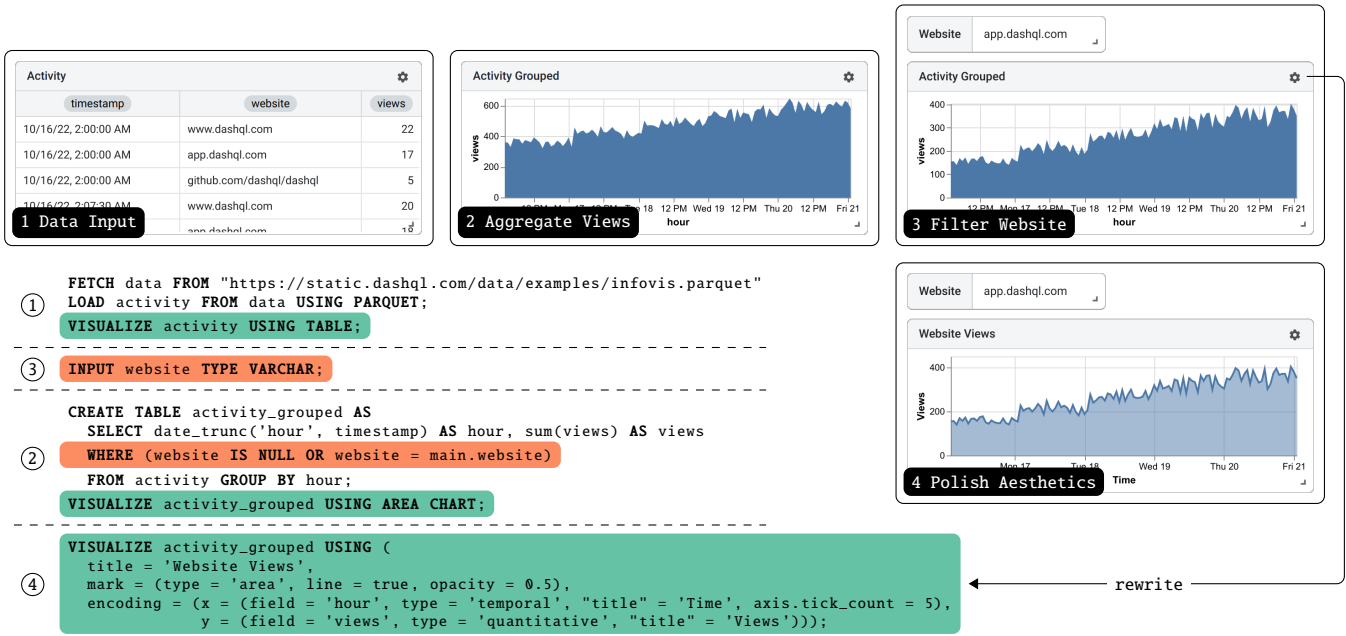}
    \vspace{0.2cm}
    \caption{
        Authoring an example analysis workflow with DashQL.
        The workflow explores website activity data in four steps.
        The steps are labeled with \Circled{1} to \Circled{4} and associate textual changes in the script with adjusted visual output.
        Visualization statements are colored in \wordbox[fill=color-visualize]{green}, the input statement and the corresponding predicate in \wordbox[fill=color-input]{orange}.
    }
    \label{f:demo}
\end{figure*}

M4 is a value-preserving time series aggregation equivalent to the one listed in \cref{f:m4}.
The query scans the relation \texttt{user\_data} and computes the four aggregates grouped by a bin key.
Afterward, the query resolves the corresponding \texttt{x}- and \texttt{y}-values of the aggregates by joining the aggregates again with the input data.
A tuple in the input qualifies in that join, if there exists an aggregate with the same key and either \texttt{x} or \texttt{y} equals an extreme value.
The query does not rely on any specific aggregation functions which makes it compatible with a wide variety of database systems.

Yet, the original version of M4 introduces a subtle but important assumption.
It scans the input relation twice and joins the extreme values to reconstruct the corresponding input tuples.
This assumes, that the extreme values are unique as the join might otherwise emit duplicates.
For example, a constant function like $f(x) = 42$ will resolve 42 as minimum and maximum \texttt{y} value of every group.
The following join will then emit the entire input relation since all tuples contain the same value for \texttt{y}.
To support non-unique \texttt{y}-values, we therefore also have to make the output distinct on \texttt{k}, \texttt{x}, and \texttt{y}.
M4 therefore consists of a repeated scan, a join and two aggregations or otherwise has to fall back to significantly slower window functions.

We propose an alternative version of M4, called AM4, shown in \cref{f:am4}.
It uses the aggregation functions \texttt{arg\_min} and \texttt{arg\_max}, sometimes implemented as \texttt{min\_by}, and \texttt{max\_by}, that are provided by several databases today (e.g., by ClickHouse, DuckDB and Presto).
The function \texttt{arg\_min(a, b)} selects an arbitrary attribute for \texttt{a} where \texttt{b} is minimal and can be computed alongside a \texttt{min(b)} aggregate at negiligble costs.
We extend M4 by additionally computing the aggregates \wordbox[fill=color-visualize]{\texttt{arg\_min(y, x)}}, \wordbox[fill=color-fetch]{\texttt{arg\_max(y, x)}}, \wordbox[fill=color-input]{\texttt{arg\_min(x, y)}} and \wordbox[fill=color-load]{\texttt{arg\_max(x, y)}}.
This resolves existing points associated with the extreme values in a single efficient grouping, eliminating the second scan and the distinct aggregation.

\subsubsection{Adaptive Materialization}

A second example for holistic optimization is called Adaptive Materialization.
DashQL statements like \texttt{FETCH} and \texttt{LOAD} only \emph{declare} data sources and formats.
It is left the optimizer to decide at runtime if the file contents should be materialized as table upfront or if the data should be loaded lazily as part of a following SQL query.
This decision not only depends on a single query but the entire script context as multiple statements might refer to the same data.
If the file format allows it, DashQL can further use projection and predicate pushdown of databases to only fetch relevant parts of a file based on the specific query.

Predicate pushdown is a common optimization technique in databases and describes the evaluation of predicates as far down in the query plan as possible.
The direction \emph{down} refers to the widespread representation of relational algebra where relations form \emph{leaves} of a tree that are combined using joins.
When optimizing relational algebra, a common task is to push individual predicates towards these \emph{leaves} to reduce the cardinality of a relation as early as possible.
If such a predicate is evaluated right after scanning file formats like Parquet, the database can evaluate the predicates on file statistics and skip reading entire row groups.

The database \texttt{DuckDB}, for example, supports reading remote Parquet files partially using a HTTP filesystem and skips row groups based on predicates in the table function \texttt{parquet\_scan}.
With DuckDB, DashQL fetches and loads the Parquet files in following SQL queries, if the data is not consumed by multiple statements.
Formats like CSV, on the other hand, require downloading and parsing the entire file, independent of subsequent filters.
In these cases, DashQL materializes the CSV contents once and shares the table with all following statements.
The decision to materialize data therefore depends on the data source, the data format, all queries in the script and the capabilities of the underlying database.
We call this technique Adaptive Materialization and see it as an opportunity to replace traditional caching logic with query-driven optimization passes.
\section{Example Data Exploration}
\label{sec:evalexplore}

We demonstrate data exploration with DashQL by constructing an example analysis workflow.
The example analyzes a dataset with website activity data and builds a dashboard to view daily total page views for individual websites.
We describe the textual changes to the script in every step and how they affect the reevaluation of the derived task graph.
The script text and the associated output of the tool are shown in \Cref{f:demo}.

\textbf{Data Input.} Our exploration begins with a declaration of the workflow's input data.
The first script is labeled with \Circled{1} and consists of three DashQL statements.
A \texttt{FETCH} statement declares that a file with name \texttt{data} can be retrieved using HTTP, a \texttt{LOAD} statement interprets this data as Parquet file and a \texttt{VISUALIZE} statement colored in \wordbox[fill=color-visualize]{green} displays the file contents.
The figure also presents the output of the first statements that visualizes the unaggregated site activity data using a single table.
This table is virtualized, which means that only visible rows are rendered.
In SQL, this virtualization translates to \texttt{LIMIT} and \texttt{OFFSET} clauses to only query the relevant subset of the data.
With a coherent language model, we can propagate the \texttt{LIMIT} and \texttt{OFFSET} specifiers towards the data retrieval during an optimization pass.
As a result, this first step only reads the file metadata and the first bytes of the Parquet file using HTTP range requests.
When the user scrolls through the data, the table dynamically reads following tuples by adjusting both specifiers.
The internal WebAssembly database also uses an accelerating readahead buffer for the remote file to minimize the number of roundtrips to the remote server.
This reduces the latency that users have to wait until seeing a visualization and provides a graceful fallback to large reads when the data is being requested.

\textbf{Aggregate Views.} Next, we want to aggregate the site activity to inspect the hourly sum of page views.
We modify the script as shown in \Circled{2} and add an explicit SQL statement that groups the site activity data as well as an additional \texttt{VISUALIZE} statement in \wordbox[fill=color-visualize]{green} to display the aggregates using an area chart.
During reevaluation, the former workflow state is left untouched since the previous statements were neither modified, nor invalidated.
The new query statement, however, needs to scan the attributes \texttt{timestamp} and \texttt{views} of all tuples in the Parquet file to compute the new aggregates.
The additional visualization statement waits for the grouping to complete and then displays an area chart.
This demonstrates the generation of Vega-Lite specifications as outlined in \Cref{sec:complementingvega} since the tool automatically selects the time and sum attributes for the x- and y-values and identifies temporal and quantitive axes.

\textbf{Filter Website.} The next step makes the analysis dashboard interactive.
Instead of showing the total page views across all websites, we want to filter the activity data by a website name that is provided dynamically by the user.
For this, \Circled{3} introduces an \texttt{INPUT} statement colored in \wordbox[fill=color-input]{orange} and includes a filter predicate in the SQL statement.
The new input with name \texttt{website} is of type \texttt{VARCHAR} and displays a text field on top of the previous area chart.
The added filter predicate checks if the website is either NULL or if the website attribute of the tuple equals the website variable in the script.
By default, the input value will be NULL which means that the dashboard will show the total page views until a website name is entered.
During reevaluation, the Patience Diff algorithm identifies the additional \texttt{WHERE} clause in the query statement and marks it as updated.
The system therefore drops and recreates the grouped activity table as well as the area chart that consumes its data.
The query now filters the attribute \texttt{website}, which means that an additional column needs to be fetched from the remote Parquet file.
This input statement shows the capability of DashQL to parameterize any SQL statement without explicit text instantiation.
The AST allows us to reference the input variable by qualifying its name with the default schema.

\textbf{Polish Aesthetics.} The last step polishes the aesthetics of the generated analysis dashboard.
The short syntax of DashQL offers a frictionless visualization of arbitrary SQL statements but may be insufficiently generic for a final workflow output.
For example, the former area chart visualization falls back to the SQL attribute names for axis labels and default colors for the covered area.
As described in \Cref{sec:complementingvega}, DashQL internally lowers the short syntax to verbose specifications.
To adjust fine-granular settings, DashQL can therefore rewrite existing statements and specify all lowered options explicitly.
\Circled{4} demonstrates this by replacing the single area chart visualization with explicit settings after interacting with the previously rendered chart.
It uses the verbose specification to adjust the title, the axis labels, the tick count and the area opacity in the workflow script.

This example demonstrates that DashQL allows for a progressive construction of analysis workflows.
The interplay between textual adjustments and continuous visualizations provides short feedback loops during the data exploration.
Propagating limit and offset specifiers is an example for a holistic optimization that reduces the amount of loaded data based on user input.

\section{Visualization with AM4}
\label{sec:evalam4}

\begin{figure}
    \includegraphics[width=\linewidth]{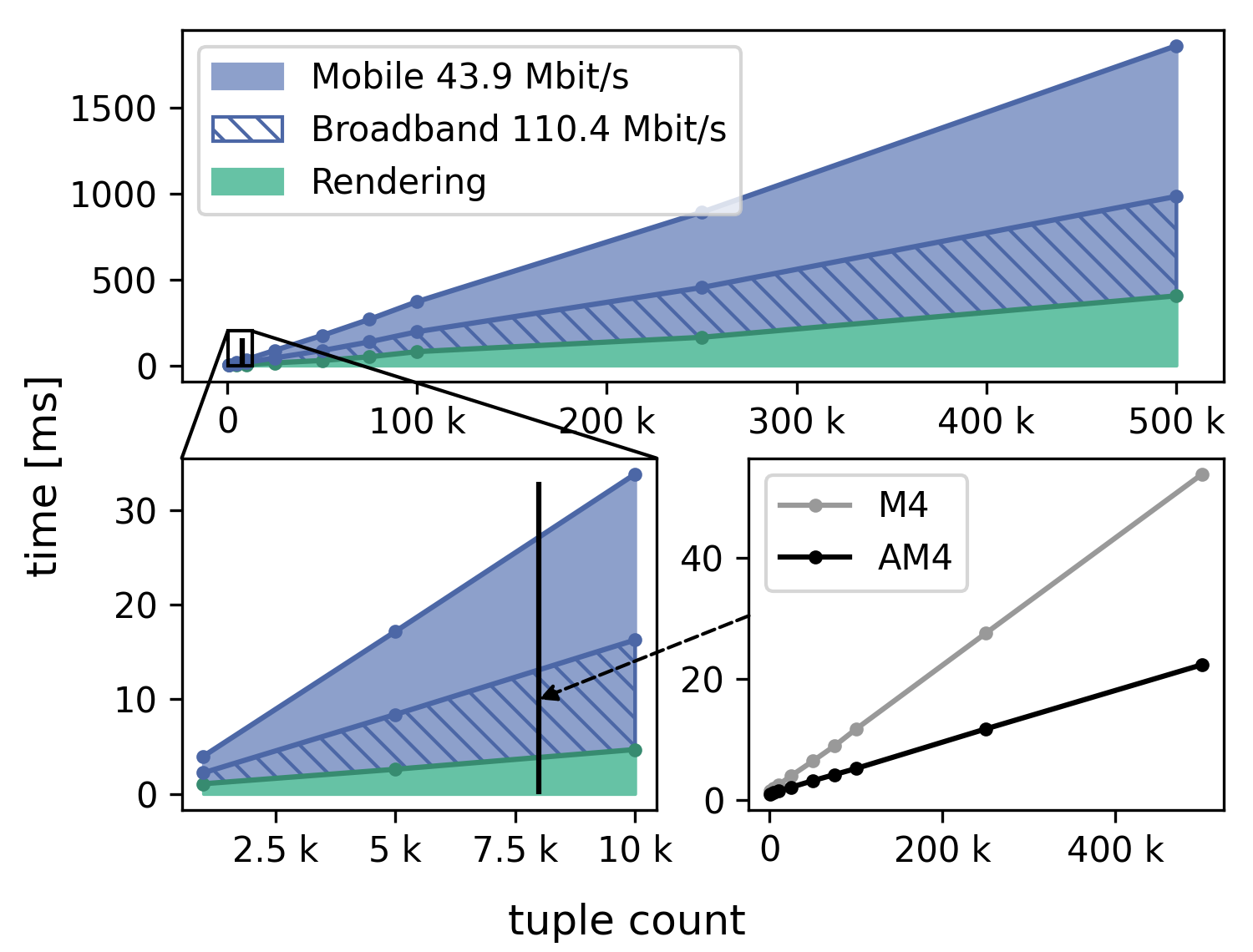}
    \caption{
        Downloading and rendering dominate the visualization times for increasing data sizes in a client-server setting.
        M4 and AM4 efficiently reduce large datasets to a small cardinality that can be visualized quickly.
    }
    \label{f:evalam4}
\end{figure}

In this section, we measure the performance of AM4, a visualization-driven aggregation and an example for a holistic optimization in DashQL.
As described in \Cref{sec:holisticopt}, AM4 accelerates chart rendering and reduces the total amount of downloaded data in a client-server setting by filtering minimum and maximum values of grouped data.
We want to demonstrate the effects of this optimization by analyzing render and download times with increasing data sizes.
The experiments were performed on a Ryzen 5800X CPU with Node.js v17.6.0 that is powered by the V8 engine v9.6.

\Cref{f:evalam4} contains three plots.
The plot at the top shows in \wordbox[fill=color-visualize]{green} color the time it takes to draw all points on a Cairo-backed canvas, using a prepared Vega view.
It further adds the time to download the data in \wordbox[fill=color-fetch]{blue} with either mobile (Cellular) or fixed broadband speeds.
The Cisco Annual Internet Report \cite{ciscoreport2020} projects average global network performances of \SI{110.4}{\mega\bit\per\second} for fixed broadband and \SI{43.9}{\mega\bit\per\second} for mobile networks by 2023.
We assume a small record size of \SI{16}{\byte} and compute the required time to download all tuples without any network latencies.
Both durations present a significant delay especially for data sizes beyond 100k tuples that hamper any interactive exploration.
The plot in the bottom right shows execution times of M4 and AM4.
If we assume a canvas width of 1000 pixels and a device pixel ratio of 2, M4 and AM4 reduce the data cardinality to 8k tuples.
AM4 computes the aggregates for a relation with 500k entries in \SI{22.3}{\milli\second} and is twice as fast as M4 which takes \SI{53.7}{\milli\second}.
The plot in the bottom left shows the render and download and times for up to 10k tuples.
A vertical line marks the resulting 8k tuples emitted by both algorithms that can be visualized quickly.

The experiment shows that both, M4 and AM4, accelerate the visualization of large data sets.
This holds even when computing the analysis locally without downloads since rendering alone becomes expensive with an increasing number of tuples.
AM4 is therefore a good example for an optimization that propagates data from visualizations, such as the canvas width, back to the SQL query.
\section{Related Work}
\label{sec:relatedwork}

DashQL builds on ideas from declarative visualization and analysis languages and automatically optimizes workflows to make them more scalable.

\subsection{Declarative Analysis Languages}

Visualization and analysis languages are fundamental to exploratory analysis, either as a programming interface or as the underlying representation of a UI tool.

Declarative, textual languages provide a high-level notation to describe data science workflows.
They often come with runtimes that optimize the data representation and query execution.
These advantages make them a popular choice over imperative languages like Python, R, or JavaScript.
For example, SQL remains a popular tool for data scientists to express queries to databases decades after its invention~\cite{chamberlin1974sequel}.
Additionally, many declarative visualization and analysis languages have emerged.
Vega~\cite{satyanarayan2015reactive} and Vega-Lite~\cite{satyanarayan2016vega}, for example, describe visualizations in JSON syntax.
Their runtimes reduce redundant computation in these specifications and fill in rendering details.
DashQL extends this research around declarative visualizations by integrating Vega-Lite specifications in the \texttt{VISUALIZE} statements.
Vega also supports declarative data loading and transformations but authoring and debugging them can be cumbersome~\cite{hoffswell2016visual} and is often not performant enough.
Dedicated analysis languages can fill this gap, for example by extracting analytical queries into explicit steps that can be annotated and tracked.
Glinda~\cite{glinda} is a declarative format for specifying data science workflows including data loading, transformation, machine learning, and visualization.
In contrast to Vega and Vega-Lite, Glinda describes analysis steps in YAML.
DashQL follows the principles of declarative analysis languages but extends the language SQL instead.
This approaches the goal of a coherent analysis language from the opposite direction as data ingestion and visualization are embedded into the database query language itself.

Vega, Glinda and DashQL, as most analysis languages, build on relational algebra and share a similar expressiveness in terms of the analyses they can describe~\cite{codd2002relational}.
Beyond the analysis steps, DashQL specifications describe inputs via UI widgets and outputs via tables and visualizations.
Unlike Precision interfaces~\cite{precisionifs} and the recent PI2~\cite{pi2} which implicitly generate UIs from SQL queries, the UI components in DashQL are explicitly described using the statements \texttt{INPUT} and \texttt{VISUALIZE}.
Like Vega visualizations, DashQL dashboards are interactive and update reactively to changes.
Vega proposed a reactive runtime for visualizations~\cite{satyanarayan2015reactive} but all declarative components need to be specified by the author.
When a declaration changes, the runtime needs to re-parse and re-evalute the entire JSON specification.

When languages are used as model in a UI tool, analysts interactively modify an underlying specification that the system can reason about~\cite{heer2015predictive}.
Polaris, which led to the creation of Tableau, explored this concept with the language VisQL~\cite{polaris}.
In Voyager~\cite{voyager}, people interactively change CompassQL~\cite{compassql} specifications and a recommender system suggests a gallery of visualizations.
Lyra is an interactive visualization design environment that authors Vega-Lite specifications on behalf of the user \cite{satyanarayan2014lyra} .
We show in the paper that DashQL extends these ideas with an compact AST representation that allows for efficient updates.

Systems often blend code and graphical interfaces and allow modifications through either direct manipulation or code.
Mage~\cite{magefluid} and B2~\cite{wu2020b2} blend the boundaries between code and UI in Jupyter Notebooks.
In Sketch-n-Sketch~\cite{sketchnsketch}, people can write a program to generate graphics or manipulate the graphics directly in the rendering canvas.
Inspired by these ideas, DashQL scripts describe visualizations like inputs, tables, and charts with text, but users can also change the statement by interacting with the UI.
For example, DashQL offers to expand the short syntax of \texttt{VISUALIZE} statements or updates chart dimensions in the text when resizing the UI widget.

\subsection{Scalable Visual Analysis}

Even small latencies in visual analysis systems negatively affect people's behavior during data exploration~\cite{liu2014effects,zgraggen2016progressive}.
Intial exposure to delays impair the subsequent performance even when delays are removed.
Therefore, we want DashQL to respond to user interactions with low latency.
DashQL builds on two ideas to achieve this goal.

First, DashQL uses an efficient in-browser analytical database based on DuckDB~\cite{duckdb}.
The database allows to evaluate analysis workflows entirely on the client, avoiding costly roundtrips to a backend server.
This lays the foundation for a distributed evaluation of workflows in the future that optimize dynamic client server scaling using a cost model~\cite{dynamicclientserver}.
Second, DashQL leverages the declarative format of analysis scripts to apply known optimizations from the database literature.
These optimizations reduce redundant and unnecessary computations and avoid loading data that is not needed to answer a query.
For example, DashQL reads data dynamically from remote files based on query predicates and projected attributes~\cite{datatilesinteractive}.
Propagating such information across statements shares similarities with provenance-supported interactions described by Psallidas et al.~\cite{provenanceinteractive}.
DashQL further implements a variant of the algorithm M4~\cite{m4} to reduce the rendering overhead with time series data.

Ideally, these optimizations happen transparently without the user having to manually specify them (as they would need to if they wrote their analysis in e.g., D3).
Previous systems~\cite{immens,nanocubes,falcon} used specialized engines to enable interactive response times.
DashQL does not yet apply some of the indexing techniques these systems proposed but it supports a wide range of analysis scenarios through general SQL queries.
VegaPlus~\cite{vegaplus} is a related project that aims to improve performance of general visualizations by extracting data transformations from Vega~\cite{satyanarayan2015reactive} specifications and running them in a system that is more scalable than the Vega runtime.
VegaPlus shifts computation but does not automatically apply data reduction techniques like M4.

% SQL~\cite{chamberlin1974sequel} is the most popular expression languages of relational algebra~\cite{codd2002relational} and is the standard langauge for expresing queries to databases.
% Polaris, ggplot2, ggvis, protovis, d3, vega
% Declarative languages are particularly well suited to represent the state of a UI tool because they are easier to reason about and optimize.

% Vega-Lite \cite{satyanarayan2016vega}
% Grammar of Graphics \cite{wilkinson2012grammar}
% B2

% Polaris \cite{polaris},
% VegaPlus \cite{vegaplus},
% Dynamic Client Server Optimization \cite{dynamicclientserver},
% DVMS \cite{dvms},
% M4 \cite{m4},
% Nanocubes \cite{nanocubes},
% Falcon \cite{falcon},
% ImMens \cite{immens},
% Sketch-n-Sketch \cite{sketchnsketch},
% Benchmarking Real-Time Interactive Querying \cite{interactivebenchmark},
% Dynamic Prefetching of Data Tiles for Interactive Visualization \cite{datatilesinteractive},
% Expressiveness of the data flow and data state models in visualization systems \cite{expressiveness},
% Interactive Visualization of Large Data Sets \cite{interactivelargedata},
% mage: Fluid Moves Between Code and Graphical Work in Computational Notebooks \cite{magefluid},
% Provenance for Interactive Visualizations \cite{provenanceinteractive},
% Demonstration of PI2: Interactive Visualization Interface Generation for SQL Analysis in Notebook \cite{pi2},
% Precision Interfaces \cite{precisionifs}

% \newpage\phantom{bla}
% \newpage

\section{Discussion}
\label{sec:summary}

This paper introduces the language DashQL.
We list example scripts throughout the sections and discuss iterative data exploration in \Cref{sec:evalexplore}.
The examples demonstrate the proximity of the language to SQL and the capability to describe complete analysis workflows.
DashQL extends SQL by defining how data can be resolved and how results should be visualized.
The coherent language model facilitates holistic optimizations covering data input, transforms and visualizations.
Nevertheless, we identify two major areas for future improvements.

First, DashQL models interactivity with the dedicated input statement.
These statements render form controls in the resulting dashboard and reevaluate parts of the task graph upon user interaction.
This enables parameterizing arbitrary SQL statements but raises the question how to support interactions with rendered visualizations.
A prominent example is cross-filtering where user interactions with one visualization translate into applied filters for another.
A possible solution could be to expose the concept of Vega-Lite signals to the language, for instance, by updating values for input statements through brushes in a time series chart.
Future versions of DashQL should therefore focus on integrating input variables as more than just constant scalar values in SQL queries.

Second, the language DashQL does not specify yet how a workflow can be executed across multiple machines.
Traditionally, there has been a clear separation between the analytics server performing computations and the client visualizing the results.
With DashQL, these boundaries are blurred as a system could spread the execution of a workflow across multiple machines.
The distributed evaluation of a workflow becomes an optimization problem that has to consider the data locality, the bandwith and computation capacities and resulting interaction latency.
For example, large datasets might require to evaluate certain predicates close to the data in the cloud but might still favor analyzing the filtered results locally.
We see DashQL as a step towards distributed analysis workflows that optimize for low interaction latencies even on large data sizes.

\acknowledgments{
This project has received funding from the European Research Council (ERC) under the European
Union's Horizon 2020 research and innovation programme (grant agreement No 725286). \hspace{1mm} \includegraphics[height=3mm]{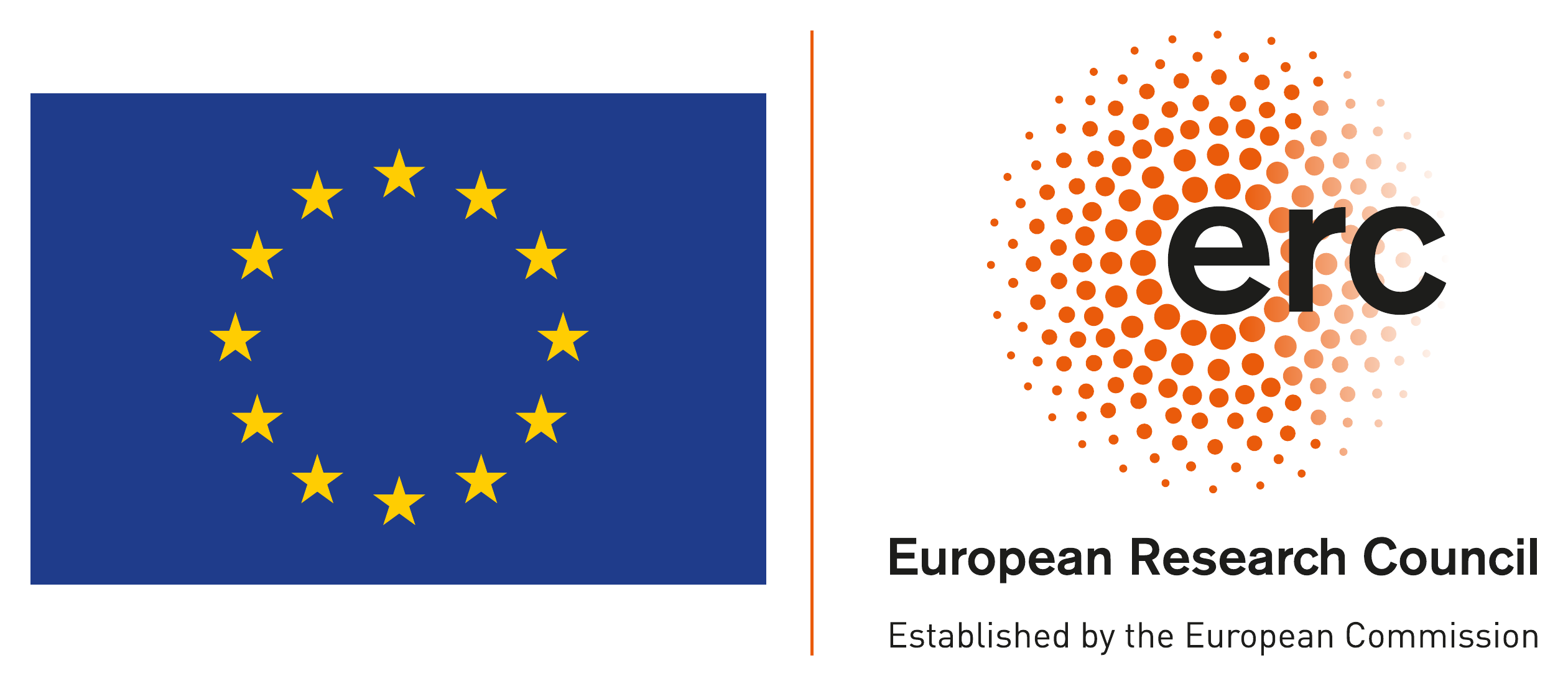}
}

%\bibliographystyle{src/theme/vgtc/abbrv-doi-hyperref-narrow}
%\bibliographystyle{src/theme/IEEEtran}
%\bibliography{main}

% -----------------------------------------------------------
% Bibliography

\end{document}